\documentclass[twocolumn]{aastex63}
\usepackage{amsmath}
\usepackage{lipsum}
\usepackage[normalem]{ulem}

\newcommand\redst{\bgroup\markoverwith{\textcolor{red}{\rule[0.5ex]{2pt}{0.4pt}}}\ULon}

\accepted{March 29, 2023}
\submitjournal{ApJ}

\shorttitle{Braking LMXBs \& Single Stellar Models}
\shortauthors{Gossage et al.}
\graphicspath{{./}{figures/}}

\begin{document}

\title{Magnetic braking with MESA evolutionary models in the single star and LMXB regimes}

\correspondingauthor{Seth Gossage}
\email{seth.gossage@northwestern.edu}

\author[0000-0001-6692-6410]{Seth Gossage}
\affiliation{Center for Interdisciplinary Exploration and Research in Astrophysics (CIERA), Northwestern University, 2145 Sheridan Road,
Evanston, IL 60208, USA}

\author[0000-0001-9236-5469]{Vicky Kalogera}
\affiliation{Center for Interdisciplinary Exploration and Research in Astrophysics (CIERA), Northwestern University, 2145 Sheridan Road,
Evanston, IL 60208, USA}

\author[0000-0001-9037-6180]{Meng Sun}
\affiliation{Center for Interdisciplinary Exploration and Research in Astrophysics (CIERA), Northwestern University, 2145 Sheridan Road,
Evanston, IL 60208, USA}




\begin{abstract}

Magnetic braking has a prominent role in driving the evolution of close low mass binary systems and heavily influences the rotation rates of low mass F- and later type stars with convective envelopes. Several possible prescriptions that describe magnetic braking in the context of 1D stellar evolution models currently exist. We test four magnetic braking prescriptions against both low mass X-ray binary orbital periods from the Milky Way and single star rotation periods observed in open clusters. We find that data favors a magnetic braking prescription that follows a rapid transition from fast to slow rotation rates, exhibits saturated (inefficient) magnetic braking below a critical Rossby number, and that is sufficiently strong to reproduce ultra compact X-ray binary systems. Of the four prescriptions tested, these conditions are satisfied by a braking prescription that incorporates the effect of high order magnetic field topology on angular momentum loss. None of the braking prescriptions tested are able to replicate the stalled spin down observed in open cluster stars aged 700 - 1000 Myr or so, with masses $\lesssim 0.8\ M_{\odot}$.

\end{abstract}



\section{INTRODUCTION} \label{sec:intro}

Magnetic braking dominates the angular momentum evolution of F- and later type (considered ``low mass'' in this work) main sequence stars that possess deep outer convective layers. Magnetic braking is a major physical mechanism by which stars in this mass range are observed (first by \citealt{kraft:1967} and \citealt{skumanich:1972}) to slow their rotation rate as they age. The theory describes how a star's surface magnetic field, coupled to its outflowing stellar wind, exerts a torque on that star \citep{schatzman:1962,weber.davis:1967,mestel:1968}, braking its rotation rate over time. Magnetic braking is the basis of gyrochronology \citep{barnes:2003,mamajek.hillenbrand:2008}, plays a key role in close, low-mass binary system evolution \citep{podsiadlowski.etal:2002, warner:2003}, and is a fundamental aspect of low-mass stellar evolution in general. In light of recent observations, an accurate model of magnetic braking (and generally stellar angular momentum transport) is still under investigation. Namely, a magnetic braking model that can simultaneously describe the recently observed stalled braking behavior \citep{curtis.etal:2019} and general spin evolution of low mass stars across time as observed in open clusters and binary systems.

A plethora of stellar rotation periods measured within the last decade from ground- and space-based missions, with many utilizing \textit{Kepler (K2)}, (e.g., \citealt{irwin.etal:2008, meibom.etal:2011, meibom.etal:2015, mcquillan.etal:2014, rebull.etal:2016,rebull.etal:2018, douglas.etal:2017, agueros.etal:2018,curtis.etal:2019}) have revealed a more detailed picture of angular momentum evolution in low-mass (in this context $\lesssim 1.3\ M_{\odot}$) stars since the time of \cite{skumanich:1972}. It now appears that stars do not steadily lose angular momentum as they age. Instead, stars in the mass range of at least about $\lesssim 0.8\ M_{\odot}$ stall between ages of at least 700 - 1000 Myr \citep{curtis.etal:2019, curtis.etal:2020, hall.etal:2021}; although stalling may occur generally in low mass stars $\lesssim 1.3\ M_{\odot}$ at different epochs (e.g., see \citealt{curtis.etal:2020, hall.etal:2021}). Solar-like stars older than the age of the Sun are also observed to rotate more rapidly than models tend to predict \citep{vansaders.etal:2016, vansaders.pinsonneault.barbieri:2019}. Both phenomena suggest that stars undergo phases of weakened or stalled magnetic braking. 

In addition to these recent findings, models have sought to explain how and why magnetic braking becomes less efficient as stellar mass decreases, where lower mass stars have been observed to spin down less efficiently than higher mass counterparts since at least the mid '80s (see \cite{krishnamurthi.etal:1997} for an early and brief review). It may seem obvious that a phenomenon involving such interrelated physical processes should vary by stellar class, but understanding why and how has remained elusive for low mass stars. Models have explored the ideas of a ``saturated'' braking state \citep{macgregor.brenner:1991, chaboyer.demarque.pinsonneault:1995, sills.etal:2000, spada.etal:2011, vansaders.pinsonneault:2013, gallet.bouvier:2013, gallet.bouvier:2015, matt.etal:2015}, core-envelope coupling \citep{bouvier:2008,irwin.bouvier:2009, denissenkov.etal:2010, spada.etal:2011, lanzafame.spada:2015, spada.lanzafame:2020}, and recent explorations of magnetic field geometry \cite{reville.etal:2015,garraffo.drake.cohen:2015,garraffo.drake.cohen:2016,garraffo.etal:2018, finley.matt:2018, finley.see.matt:2019,see.etal:2019, see.etal:2020} and differential rotation \cite{ireland.etal:2022}. 

All of these theoretical innovations have been linked to certain observations. Magnetic field saturation has followed from the observation that magnetic field activity saturates below a critical Rossby number\footnote{I.e., $R_{\rm o} = (\omega \tau_c)^{-1}$, decreases both as rotation rate ($\omega$) and convective turnover time ($\tau_c$) rise. The convective turnover time, written $\tau_c = H_P / v_c$ (with $H_P$ as the pressure scale height, $v_c$ the convective velocity), typically rises with decreasing stellar mass/increasing convection zone depth (e.g., see \citealt{cranmer.saar:2011}).} \citep{noyes.etal:1984,pizzolato.etal:2003, wright.etal:2011, vidotto.etal:2014, wright.drake:2016, newton.etal:2017, wright.etal:2018,see.etal:2019b, medina.etal:2020}. Some stars (including the Sun, depending on solar minimum or maximum) are suggested to have magnetic field topologies that are more complex than purely dipolar via Zeeman-Doppler imagining (ZDI) magnetograms \citep{donati.landstreet:2009, marsden.etal:2011, waite.etal:2015}. As is also observed in the Sun, many low mass stars have been observed to exhibit differential rotation, the effects of which have been recently explored by \cite{ireland.etal:2022}. Meanwhile, core-envelope coupling, though not directly observed, is currently the only model \citep{spada.lanzafame:2020} capable of replicating the stalled magnetic braking observed in open clusters aged roughly 700 - 1000 Myr \citep{curtis.etal:2020,hall.etal:2021}.

Besides analytical (and semi-empirical, e.g., \citealt{barnes:2003, barnes:2010, angus.etal:2015, angus.etal:2019, kounkel.etal:2022}) models, 1D stellar evolution models remain the only tractable method to simulate and examine entire stars over a stellar lifetime. However, 1D stellar evolution models cannot truly simulate multidimensional phenomena like convection or magnetic fields, and so rely on prescriptions to describe how they scale with stellar properties like mass, pressure, temperature, etc. (mixing length theory -- MLT -- for convection \citealt{bohm-vitense:1958, henyey.etal:1965, cox.giuli:1968} and re-parameterizations of the Alfv\'en radius for stellar magnetic fields are examples). Relatively recent work has used 2D \citep{matt.pudritz:2005,matt.pudritz:2008a,matt.pudritz:2008b,matt.etal:2012,matt.etal:2012b}, 2.5D \citep{reville.etal:2015, finley.matt:2017,finley.matt:2018}, and 3D \citep{garraffo.drake.cohen:2015, garraffo.drake.cohen:2016} magnetohydrodynamic (MHD) simulations to investigate how wind and magnetic field properties scale with mass, rotation rate, and thermodynamic quantities.

For instance, while \cite{matt.etal:2015} present a generalized prescription (based on \citealt{kawaler:1988}) that replicates the spin down of late F-, G-, and K-type stars quite well, it struggles to represent M dwarf spin down and the bimodal morphology of open cluster rotation period data (represented by the ``C'' and ``I'', or fast and slow branches named by \citealt{barnes:2003}). \cite{brown:2014} presented a prescription replicates this bimodal morphology called the ``metastable dynamo''; however, it was unclear what physical mechanism might drive this behavior. \cite{garraffo.drake.cohen:2016,garraffo.etal:2018} have provided a prescription that codifies the effects of complex multipolar magnetic fields (i.e., quadrupolar, octopolar, etc. and hereafter, simply ``complexity'') from their MHD simulations that can also successfully capture the observed fast and slow branch morphology \citep{garraffo.etal:2018,gossage.etal:2021}. Some recent simulations question whether higher order modes have a significant influence on the stellar magnetic field structure (e.g., \citealt{finley.matt:2018,finley.matt.see:2018,see.etal:2019,see.etal:2019b,finley.see.matt:2019,see.etal:2020}). These recent developments have brought forth deeper questions about the stellar magnetic field, wind, and dynamo driving magnetic braking.

The previous studies have mainly been applied to the study of single stars, which provide a relatively simple scenario to study magnetic braking in. However, the evolution of binary star systems can be heavily influenced by this phenomenon as well, especially for close binaries such as low-mass X-ray binaries and cataclysmic variables  (LMXBs and CVs; e.g., \citealt{kalogera.webbink:1996,ivanova.taam:2003}). Magnetic braking is a primary means of angular momentum extraction for these systems, shrinking their orbital separation, and driving mass transfer between stellar companions. In applying magnetic braking prescriptions to binary star models, these systems are always assumed to be tidally locked, typically following the prescription of \cite{rappaport.verbunt.joss:1983}. In close orbits, companions are generally expected to be tidally locked (at least for those with orbital periods less than ~8 days \citealt{mazeh:2008}). This means that angular momentum extracted from one companion may also be extracted from the system's orbital angular momentum \citep{verbunt.zwaan:1981}. 

\cite{rappaport.verbunt.joss:1983} presented a prescription based on this and the Skumanich law that is prominently used in binary evolution studies still today. \cite{van.ivanova:2019b} presented a prescription based on \cite{mestel.spruit:1987} that replicates LMXB data where other prescriptions have issues. \cite{deng.etal:2021} found this to be the case as well in a comparative analysis with several magnetic braking prescriptions, and the prescription of \cite{van.ivanova:2019b} has recently been adopted in e.g., \texttt{DABS}, a database of LMXB models \citep{singh-mangat.etal:2022}. However, \cite{el-badry.etal:2022} have recently shown that CV data seem to favor models including a ``saturation'' or weakened magnetic braking effect, as in \cite{matt.etal:2015} or \cite{garraffo.etal:2018}, as was also found by \cite{andronov.pinsonneault.sills:2003}. While \cite{rappaport.verbunt.joss:1983} and CARB \citep{van.ivanova:2019b} magnetic braking have been shown to successfully model binary systems, they less commonly been applied and studied in the context of single stellar evolution.

In this study, we aim to compare and constrain four magnetic braking prescriptions: M15 \citep{matt.etal:2015}, G18 \citep{garraffo.etal:2018}, CARB \citep{van.ivanova:2019b}, and RVJ83 \cite{rappaport.verbunt.joss:1983} implemented in 1D stellar evolution models to recreate \textit{both} single star open cluster rotation period and LMXB rotation period data in working towards a consensus prescription. Our motivation for including these four prescriptions in particular is due to the prevalence of both \cite{rappaport.verbunt.joss:1983} and \cite{matt.etal:2015}; the basis on relatively recent MHD simulations for both the \cite{matt.etal:2015} and \cite{garraffo.etal:2018} prescriptions; and the recent proposal of CARB as an alternative to \cite{rappaport.verbunt.joss:1983} by \cite{van.ivanova:2019b} in binary stellar evolution models.

In the following sections, we first describe our single star open cluster and LMXB data sets, and provide their sources (Sec. \ref{sec:data}). We then describe our 1D stellar evolution models, and describe the four magnetic braking prescriptions that have been incorporated (Sec. \ref{sec:models}). Our results, comparing these models to data, are presented in Sec. \ref{sec:results}. We then discuss our results and conclude in Sec. \ref{sec:discussion.conclusions}.

\section{DATA DESCRIPTION}\label{sec:data}

We compare our single star models to stellar rotation periods (obtained by others through light curve analyses) and masses (estimated by others originally through empirical color-mass relations) observed in open clusters. These clusters range in age from roughly 2 - 2500 Myr, tracing the rotation period evolution of low mass stars through time. In order of age, we have used data from: NGC 6530 (2 Myr, \citealt{henderson.etal:2012}), NGC 2264 (3 Myr, \citealt{affer.etal:2013}), NGC 2362 (8 Myr, \citealt{irwin.etal:2008}, Upper Scorpius (Usco; 10 Myr, \citealt{rebull.etal:2018}), h Persei (h Per; 13 Myr, \citealt{moraux.etal:2013}), the Pleiades (125 Myr, \citealt{rebull.etal:2016}), M34 (220 Myr, \citealt{meibom.etal:2011}), M37 (500 Myr, \citealt{nunez.etal:2015}), the Praesepe (676 Myr, \citealt{douglas.etal:2017}), NGC 6811 (1 Gyr, \citealt{curtis.etal:2019}), and NGC 6819 (2.5 Gyr, \citealt{meibom.etal:2015}).

Some of these data sets likely contain binaries, but efforts were often made to identify candidates in the original studies. After accounting for binary candidates, the bulk of these data appear dominated by single stars that form the major morphological features of the data. In analysis of the Praesepe by \cite{douglas.etal:2017}, some binary candidates were found interspersed among single stars, but were often found to have more rapid rotation than their similar mass, single star counterparts; \cite{douglas.etal:2017} show them residing in the sparsely populated space between the slow- and fast-rotator branches. Analysis of USco by \cite{rebull.etal:2018} show similarly that binary candidates are interspersed with single stars, but e.g., \cite{somers.etal:2017} show that their removal does not affect the major morphology of the period-mass data. \cite{moraux.etal:2013} show binary candidates interspersed, that also tend to favor more rapid rotation rates. As demonstrated by e.g., \cite{packet:1981,demink.etal:2013}, theoretical predictions (based on e.g., calculations from \citealt{lubow.shu:1975}) also expect that binary interactions should tend to create rapidly rotating stellar products. With respect to these data, it is also important to bear in mind that masses have been inferred from photometric colors via various techniques, introducing systematic uncertainties in the estimated masses (e.g., as discussed by \citealt{breimann.etal:2021}). Thus, cited mass ranges should be considered approximate, and have been labeled as such.

We compare our binary star models to the observed periods and estimated donor masses of neutron star binaries collected in \cite{van.ivanova:2019a} (their Table 3, with references therein). These data come from Milky Way neutron star LMXBs and include ultra-compact X-ray binaries, or UCXBs (rotation periods less than 80 minutes) where the donor star may be degenerate. Our binary star models assume a representative neutron star mass of 1.4 $\rm M_{\odot}$, in accordance with the modelling of \cite{van.ivanova:2019b}. In reality, neutron star masses may range as high as about 2 $\rm M_{\odot}$ \citep{antoniadis.etal:2013,rezolla.most.weih:2018}.

\section{MODEL DESCRIPTION} \label{sec:models}

Our models were created with \texttt{MESA r11701} (released alongside \citealt{paxton.etal:2019}), a 1D stellar structure and evolution code. These models require certain input physics; our adopted opacities, stellar wind schemes, chemical abundances, nuclear reaction rates, boundary conditions, and MLT parameters are those used in \texttt{POSYDON} \citep{posydon.i:2022}. The \texttt{POSYDON} project adopts many of the same micro- and macrophysical assumptions made in the single star model database, \texttt{MIST} \citep{choi.etal:2016}, however it also includes input physics relevant to binary stellar evolution. For other default physics included in our \texttt{MESA} models, see the series of development papers \citep{paxton.etal:2011,paxton.etal:2013,paxton.etal:2015,paxton.etal:2018,paxton.etal:2019,jermyn.etal:2022}, with the handling of rotation described in both \cite{paxton.etal:2013,paxton.etal:2019}, and the handling of binary stellar evolution in \cite{paxton.etal:2015}. Below we describe further details relevant to the single  and binary star models. All models -- single and binary -- are given initial solar metallicity (i.e., the proto-solar values of $Z$ = 0.0142, $Y$ = 0.2703 of \citealt{asplund.etal:2009}).

\subsection{Single Star Initial Conditions} \label{ssec:ss.initial.cond}
Our single star model grid covers the mass range $\rm M = 0.1$ to $1.3\ \rm M_{\odot}$, in steps of 0.1 $\rm M_{\odot}$ and are evolved from the pre-main sequence (pre-MS) with initial rotation rates of from $(\rm \Omega/\Omega_{crit})_i = 0.02,\ 0.04,\ 0.08$ and $0.15$. Here (with $G$, $M$, $R$, and $L$ as the gravitational constant, stellar mass, radius, and luminosity),

\begin{equation}\label{eq:omega.omegacrit}
    \frac{\Omega}{\Omega_{\rm crit}} = \frac{\Omega}{\sqrt{(1-L/L_{\rm Edd})\frac{GM}{R^3}}}
\end{equation}
is the ratio of stellar surface angular velocity, $\Omega$, to critical angular velocity, $\Omega_{crit}=\sqrt{(1-L/L_{\rm Edd})\frac{GM}{R^3}}$ \citep{paxton.etal:2013}. The critical angular velocity is the angular velocity at which centrifugal forces balance gravitational forces (modified by the Eddington factor with $L_{\rm Edd}$ being the Eddington luminosity of the star); for $\Omega > \Omega_{\rm crit}$, centrifugal forces exceed gravitational, such that surface material may escape the star. We quantify our rotation rates by $\Omega/\Omega_c$ (rather than initial rotation period, $\rm P_{rot,i}$) to be consistent with large stellar model grids that use the same quantity\footnote{This is a slight difference from \cite{gossage.etal:2021} where $\rm P_{rot,i}$ was used instead. Models in the current study with the same $(\rm \Omega/\Omega_{crit})_i$ will have slightly different corresponding $\rm P_{rot,i}$ depending on their mass, but are similar.}, such as \texttt{Geneva} \citep{ekstrom.etal:2012} and \texttt{MIST} \citep{choi.etal:2016}. This range roughly corresponds to $\rm P_{rot,i} = 1.5$ to 12 days.

This range of initial rotation rates was chosen, in part, to avoid reaching super-critical rotation rates (via pre-MS contraction) prior to the zero age main sequence (ZAMS). If the initial rotation rate is too high, stars will reach super-critical rotation rates during pre-MS contraction and experience rotationally enhanced mass loss \citep{paxton.etal:2013}. This is a possibility in reality, but we avoid it as a matter of simplification. We have found that $\rm (\Omega/\Omega_{crit})_i = 0.15$ is roughly the fastest rotation rate possible before pre-MS super-critical rotation rates occur in our models near solar mass. This rotation rate limit increases towards lower masses, but we impose the solar mass limit (i.e., a maximum of $\rm (\Omega/\Omega_{crit})_i = 0.15$) on all masses for simplicity. Outside of this consideration, we also chose our rotation rate range to roughly cover the span observed for stars in young open clusters like NGC 6530 (2 Myr, \citealt{henderson.etal:2012}) and stars in h Persei (13 Myr, \citealt{moraux.etal:2013}). See also data (and their references) used in \cite{gossage.etal:2021}, from which these conditions are derived. The data itself may likely contain stars that have evolved through super-critical rotation, but we find that our models are capable of covering the bulk of the data, and models required to explain more rapid rotation rates appear to comprise a relatively small portion of the full dataset. For example, see Fig. \ref{f:all.ocs}, panel (a), where models rotating up to $\rm (\Omega/\Omega_{crit})_i = 0.15$ comprise ~70\% of the \cite{moraux.etal:2013} data, and the majority of the data at later ages e.g., as in Fig. \ref{f:all.ocs}, panels (b) - (d).

Our single star models include a basic accounting for disk locking, such that they are given an initial rotation rate, and locked to that rate until the end of their disk locking time ($\tau_{\rm DL}$). Disk locking is a phenomenon whereby T Tauri stars may strongly couple to their proto-stellar accrection disks via magnetic fields, such that angular momentum gained through accretion may balance with that lost via winds so that stars maintain a ``locked'', constant rotation rate (see e.g., \citealt{koenigl:1991,barnes.etal:2001}). After some time called the often called the ``disk locking time'', this accretion disk dissipates and the stars spin up during pre-MS gravitational collapse. We use a disk locking time of $\tau_{\rm DL} = 3$ Myr. Based on the analysis of \cite{gallet.bouvier:2013}, the disk locking time for solar-like stars may reside between 2.5-5 Myr, depending on the initial rotation period. However, disk locking itself is complex and not well understood; its time of relevance may vary with stellar initial rotation period, metallicity, and mass -- see e.g., \cite{gallet.zanni.amard:2019, pantolmos.etal:2020, roquette.etal:2021} and \cite{ireland.etal:2021} for recent investigations. We have tested varying the disk locking time in the range 2.5-5 Myr and found no major change in our results. However, other stellar model sets, such as those of \cite{amard.etal:2019} include a variable disk locking time scheme. We have opted to lock to a constant $\Omega/\Omega_{c}$ value, while real stars appear to lock to a constant $\rm P_{rot}$ value. We have tested model behavior where the rotation rate is locked to a constant $\rm P_{rot}$ instead, and found that it makes no major difference in the evolution.

\subsection{Binary Star Initial Conditions} \label{ssec:bs.initial.cond}
We computed a grid of binary star models for each of our four implemented magnetic braking prescriptions. Our binary model grid is based on the donor mass and initial rotation period ranges used by \cite{van.ivanova:2019b} to replicate the same data, with slight modifications. Donor masses range from $\rm M_{d} =  1$ to $3\ \rm{M_{\odot}}$ (in steps of 0.1 $\rm M_{\odot}$ for $\rm M_d < 2.4$ $\rm M_{\odot}$, otherwise 0.2 $\rm M_{\odot}$), as masses greater than this tend to immediately enter a common envelope and experience unstable mass transfer, which is outside of \texttt{MESA}'s current modelling capabilities (but see work by \citealt{marcahnt.etal:2021}). Models are given initial orbital periods ranging from $\rm log_{10}(P_{\rm rot}/day)_i = -0.4$ to 3.2 (in steps of 0.05), as in our modelling orbital periods much slower than this begin too widely separated to inspiral and can not replicate the data. These models use a neutron star mass of $1.4\ \rm{M_{\odot}}$, represented by a point mass that includes no internal physics whose companion (donor) is modeled as a hydrogen-burning main sequence star. Mass transfer between the donor and companion is conservative, except when the transfer rate becomes super-Eddington (as prescribed in \citealt{posydon.i:2022}).

Similar to \cite{van.ivanova:2019b}, we do not include disk locking with our binary star models as \texttt{POSYDON} evolves models from the ZAMS, based on pre-computed models evolved without the effect. These binary star models (the donor star in particular) are otherwise subject to the same treatment of rotation that our single star models experience. These binary star models terminate upon encountering unstable mass transfer (generally during a common envelope), core hydrogen depletion (of the donor), or else the age of a Hubble time. In this work, we show their evolution through the period of Roche lobe overflow (RLOF) mass transfer that occurs as orbital separation is reduced as the donor star loses angular momentum, which is driven largely by magnetic braking (e.g., \citealt{kalogera.webbink:1996,ivanova.taam:2003}).

\subsection{Magnetic Braking Prescriptions}

In this section, we briefly detail the four magnetic braking prescriptions under consideration. The four prescriptions may be placed in two classifications: those based on the empirical Skumanich law \citep{skumanich:1972} and those based on the derivation of \cite{weber.davis:1967} for angular momentum loss due to a magnetic stellar wind\footnote{The Skumanich law may emerge as a special case of the derivation of \cite{weber.davis:1967} under the assumption of a thermal wind (i.e., with wind speed proportional to the sound speed) and a radial magnetic field \citep{mestel:1984}.}. The prescriptions from \cite{rappaport.verbunt.joss:1983} and \cite{garraffo.etal:2018} fall into the former category, while those from \cite{matt.etal:2015} and \cite{van.ivanova:2019b} fall into the latter. Furthermore, \cite{matt.etal:2015} and \cite{garraffo.etal:2018b} involve 2D or 3D MHD simulations (which operate on first principles by solving MHD equations).

A major point at which each prescription begins to deviate from first principles is in how they choose to parameterize the Alfv\'en radius, $r_A$. In the context of magnetic braking, $r_A$ is sometimes colloquially referred to as the ``co-rotation radius'' or the ``lever arm'' (e.g., see \citealt{matt.pudritz:2008a, reville.etal:2015}). The Alfv\'en radius is the distance out to which magnetic energy density dominates in the wind, and beyond which kinetic energy density dominates. The Alfv\'en radius is commonly conceptualized as the distance out to which the wind co-rotates with the star, although in reality the magnetic field exerts a torque on the wind that does not abruptly lose influence past the the Alfv\'en radius. Essentially all magnetic braking formulations must parameterize $r_A$ in terms of simpler stellar properties if they are to be used in 1D stellar evolution models, as it is a property of the magnetic field structure, which can not be captured accurately via 1D stellar evolution. To this end, simple assumptions about the magnetic field and/or wind properties are typically introduced.

\cite{rappaport.verbunt.joss:1983} derive their prescription from the Skumanich law (which amounts to the relatively simple assumption of a radial field and thermal wind \citealt{mestel:1984} in the context of equations from \citealt{weber.davis:1967}), and introduce normalization factors and scaling parameters to allow flexibility. \cite{matt.etal:2015} meanwhile (similar to \citealt{kawaler:1988}) work from first principles as laid out by \cite{weber.davis:1967}, deriving a generalized prescription whose $r_A$ is parameterized and fit according to 2D MHD simulation results, assuming dipolar magnetic fields in a series of papers \citep{matt.pudritz:2008a,matt.pudritz:2008b,matt.etal:2012}. \cite{garraffo.etal:2018} allowed for more complex field geometries in 3D MHD simulations \citep{garraffo.drake.cohen:2015} that solve for angular momentum and mass loss at the Alfv\'en surface (i.e., at the surface of the volume enclosed by $r_A$) using equations from \cite{weber.davis:1967} and \cite{mestel:1999}; \cite{garraffo.etal:2018} derived their prescription from the simulation predicted dipolar loss rate, scaled by a function of the magnetic complexity fit to their simulation results in \cite{garraffo.drake.cohen:2016}. Lastly, \cite{van.ivanova:2019b} derive their prescription starting from \cite{weber.davis:1967,mestel.spruit:1987} and assume a radial field, utilizing re-parameterizations for the wind escape speed at $r_A$ from MHD simulations performed by \cite{reville.etal:2015}.

Aside from $r_A$, all prescriptions under consideration (except that of \citealt{rappaport.verbunt.joss:1983}) involve the Rossby number, $R_{\rm o}$. Dynamo models utilize $R_{\rm o}$ (e.g., \citealt{moffatt:1978,parker:1979}), where it is related to the dynamo number $N_D\approx R_{\rm o}^{-2}$ which represents the ratio of magnetic field generation to dissipation in the convection zone \citep{noyes.etal:1984}\footnote{Generally, the definition of the Rossby number goes back to the Navier-Stokes equations and its exact mathematical expression depends on the phenomenon it describes, but it is the ratio of the inertial force to the Coriolis force. In the context of stellar dynamos, it can be shown to scale as $R_{\rm o}\sim \rm{P_{rot}}/\tau_c$ \citep{moffatt:1978,parker:1979,montesinos.etal:2001}.}. A lower Rossby number is thought to correlate to more magnetic field activity under this theory, which is corroborated by observations, except that magnetic field activity appears to saturate below some critical Rossby number, i.e., for some $R_{\rm o} < R_{\rm o,sat}$ (\citealt{noyes.etal:1984,pizzolato.etal:2003,wright.etal:2011,vidotto.etal:2014,wright.drake:2016,newton.etal:2017,wright.etal:2018,see.etal:2019b, medina.etal:2020}). The Rossby number is typically expressed as $R_{\rm o} = \rm{P_{rot}}/\tau_c$, where $\tau_c$ is the convective turnover time, and $\rm{P_{rot}}$ is the stellar rotation period. The Rossby number is often used in parameterizing magnetic field properties, such as its strength \citep{matt.etal:2015,van.ivanova:2019b} or complexity \citep{garraffo.etal:2018} in magnetic braking prescriptions. Physically, the Rossby number represents some measure of how rotation and convection conspire to generate magnetic fields in dynamo theory; both large rotation rates (small $\rm P_{rot}$) and deep convection zones (large $\tau_c$) are conducive to small $R_{\rm o}$ and greater magnetic field activity (apparently up to some saturated value) under this theory.

\subsubsection{\cite{rappaport.verbunt.joss:1983} braking}
Working under the assumption of tidal locking, \cite{rappaport.verbunt.joss:1983} developed a magnetic braking formalism that is applicable to binary star systems. This is also the default magnetic braking model included in \texttt{MESAbinary} currently (implemented in \citealt{paxton.etal:2015}). \cite{rappaport.verbunt.joss:1983} infer a braking law from the \cite{skumanich:1972} relation, as in \cite{verbunt.zwaan:1981}, expressed as:

\begin{equation} \label{eq:rvj83.jdot}
 \dot{J} = -3.8\times10^{-30}M R_{\odot}^4 \left(\frac{R}{R_{\odot}}\right)^{\gamma_{\rm mb}} \Omega^3
\end{equation}

The parameter $\gamma_{\rm mb}$ is free to vary, affecting how strongly this torque scales with stellar radius. In Eq. \ref{eq:rvj83.jdot}, the stellar angular surface rotation rate ($\Omega$) is written, as in \cite{rappaport.verbunt.joss:1983}, however the orbital angular rotation rate could be used equivalently under the assumption of tidal locking, as is done in \cite{paxton.etal:2015}. \cite{rappaport.verbunt.joss:1983} did not test the dependence of this torque on stellar mass or angular rotation rate. \texttt{MESA} sets $\gamma_{\rm mb} = 3$ by default \citep{paxton.etal:2015}, but we take $\gamma_{\rm mb} = 4$ to be consistent with comparisons made by \cite{van.ivanova:2019a} in studying LMXBs who cite that $\gamma_{\rm mb} = 4$ is widely used and consistent with \cite{verbunt.zwaan:1981}. We did test model predictions at $\gamma_{\rm mb} = 3$ as well, and found no significant change in our conclusions. In principle, this prescription could be applied to both the primary and secondary star of a binary system, leading to contributions from both; as our models represent neutron star LMXBs, we only apply this prescription to the donor star.

\subsubsection{\cite{matt.etal:2015} braking}\label{ssec:m15form}

\cite{matt.etal:2015} follows from the suite of 2D MHD simulations and parameter studies performed by \cite{matt.pudritz:2005,matt.pudritz:2008a,matt.pudritz:2008b}, and \cite{matt.etal:2012, matt.etal:2012b} to generalize the torque formalism of \cite{kawaler:1988}, which itself follows from \cite{weber.davis:1967}. Differing from their previous formulations, \cite{matt.etal:2015} interwove the observed effect that magnetic activity appears to saturate below a critical Rossby number (i.e., activity saturates when $R_{\rm o} < R_{\rm o,sat}$; \citealt{noyes.etal:1984,pizzolato.etal:2003,wright.etal:2011, wright.etal:2018,wright.drake:2016,newton.etal:2017,wright.etal:2018,medina.etal:2020}). In \cite{matt.etal:2015}, this critical $R_{\rm o}$ is characterized through the free parameter $\chi=R_{\rm o,\odot}/R_{\rm o,sat}$, which appears in the equations below. This torque thus has a saturated and unsaturated regime, and is implemented in this work, as described by \cite{matt.etal:2015}, as

\begin{equation} \label{eq:m15.jdot.unsat}
 \dot{J} = -\mathcal{T}_0\left(\frac{\tau_c}{\tau_{c,\odot}}\right)^p \left(\frac{\Omega}{\Omega_{\odot}}\right)^{p+1},\ \rm{unsaturated}  
\end{equation}
in the unsaturated regime (i.e., $R_{\rm o}>R_{\rm o,sat}$), and
\begin{equation} \label{eq:m15.jdot.sat}
 \dot{J} = -\mathcal{T}_0\chi^p \left(\frac{\Omega}{\Omega_{\odot}}\right),\ \rm{saturated}  
\end{equation}
in the saturated regime, where the scaling with $R_{\rm o}$ is replaced by the constant $\chi$, leading to less efficient braking. In both cases,
\begin{equation} \label{eq:m15.jdot.const}
 \mathcal{T}_0 = K\left(\frac{R}{R_{\odot}}\right)^{3.1} \left(\frac{M}{\rm M_{\odot}}\right)^{0.5} \gamma^{-2m}    
\end{equation}
and $\gamma = \sqrt{1+(u/0.072)^2}$ from Eq. (8) of \cite{matt.etal:2012} and appendix A of \cite{reville.etal:2015}. Here  $u$ is the ratio of rotation velocity to critical rotation velocity, $v/v_{\rm{crit}}$. The constants $K$, $m$, $p$, and $\chi$ are free calibration parameters. The values that we have adopted for these calibration parameters are shown in Table \ref{tab:m15.params}, and were calibrated on open cluster data with single star models in \cite{gossage.etal:2021}. The adopted solar parameter values appearing in Eqs. \ref{eq:m15.jdot.unsat} - \ref{eq:m15.jdot.const} are the same as in \cite{matt.etal:2015, matt.etal:2019}.

\begin{table}[h!]
    \centering
    \caption{Adopted calibration parameters in our implementation of the \cite{matt.etal:2015} braking model. The corresponding values cited by \cite{matt.etal:2015}, and as revised by \cite{matt.etal:2019} appear in the second column for reference.}
    \begin{tabular}{||c|c|c||}
         Parameter &  This Work & \cite{matt.etal:2015} \\
         \hline
        K & $1.4\times 10^{30}\ \rm{erg}$ & $6.3\times 10^{30}\ \rm{erg}$ \\
        m & $0.22$ & $0.22$ \\
        p & $2.6$ & $2$ \\
        $\chi$ & $14$ & $10$ \\
    \end{tabular}
    \label{tab:m15.params}
\end{table}

\subsubsection{\cite{garraffo.etal:2018} braking}\label{ssec:g18.form}

\cite{garraffo.drake.cohen:2015,garraffo.drake.cohen:2016, garraffo.etal:2018} made use of 3D MHD simulations (see also 2.5D MHD simulations by \citealt{reville.etal:2015}) to examine the effects of complex magnetic field geometries (e.g., quadrupolar, octopolar, etc., rather than just dipolar) on mass and angular momentum loss. A major finding from these simulations (and found by \citealt{reville.etal:2015} and preliminarily by \citealt{matt.pudritz:2008a}) was that magnetic field complexity can significantly reduce mass loss rates, consequently reducing angular momentum lost to winds (i.e., magnetic braking). \cite{garraffo.drake.cohen:2016,garraffo.etal:2018} express their formulation of torque as 
\begin{equation} \label{eq:g18.jdot}
 \dot{J} = \dot{J}_{\rm dip} Q_{J}(n)   
\end{equation}
with
\begin{equation} \label{eq:g18.jdip}
 \dot{J}_{\rm dip} = - c \Omega^3 \tau_c   
\end{equation}
representing dipolar magnetic field spin down on the slow branch with a Skumanich-like form. A Skumanich-like form was chosen in \cite{garraffo.etal:2018} as a simple spin down model, but presumably any torque that accurately describes dipolar field spin down could replace Eq. \ref{eq:g18.jdip}. Here, $c$ is a free parameter setting the overall strength of $\dot{J}_{\rm dip}$.

The function $Q_J (n)$ was derived by \cite{garraffo.drake.cohen:2016} via fits to their 3D MHD simulation results and serves to modulate the dipolar angular momentum loss as a function of the magnetic field complexity, which is parameterized via $n$. This function is written as
\begin{equation} \label{eq:g18.qj}
 Q_{J}(n) = 4.05 e^{-1.4n} + \frac{n-1}{60 B n}
\end{equation}
where $B$ is the magnetic field strength. This second term involving $B$ only dominates around where $n>7$, where loss rates being to saturate, as can be seen in \cite{garraffo.drake.cohen:2016}. We ignore this term, as described in \cite{gossage.etal:2021} so that $Q_J (n)$ takes the form
\begin{equation} \label{eq:g18.qjsimple}
 Q_{J}(n) = 4.05 e^{-1.4n}.
\end{equation}
The magnetic complexity number, $n$, is defined in terms of $R_{\rm o}$, with $P$ being the rotation period of the star, $P = 2\pi/\Omega$,
\begin{equation} \label{eq:g18.n}
 n = \frac{a}{R_{\rm o}} + b R_{\rm o} + 1.   
\end{equation}
In Eq. \ref{eq:g18.n}, $n=1$ corresponds to a dipole field, while higher $n$ represents higher order magnetic fields. The parameters $a$ and $b$ are calibration constants.

Eq. \ref{eq:g18.n} was constructed in \cite{garraffo.etal:2018} to match two empirically based regimes. Firstly, at low $R_{\rm o}$, observations suggest magnetic field activity saturates, as described in Sec. \ref{ssec:m15form}, leading to inefficient magnetic braking (in the context of \cite{garraffo.etal:2018}, described by high magnetic field complexity, and thus inefficient angular momentum loss at low $R_{\rm o}$). Secondly, \textit{Kepler} observations of stars at $R_{\rm o}$ greater than about $1-2$ \citep{vansaders.etal:2016,vansaders.pinsonneault.barbieri:2019}, appear to show signs of inefficient braking at late ages and slow rotation periods (suggesting stars may enter another phase of high magnetic field complexity at high $R_{\rm o}$). Magnetic field complexity trends based on Zeeman-Doppler-Imaging \citep{donati.landstreet:2009,marsden.etal:2011,alvarado-gomez.etal:2015,waite.etal:2015} may support this scenario, but it remains unclear. The values adopted for the calibration constants $a$, $b$, and $c$ are collected in Table \ref{tab:g18.params} and were obtained through calibration on open cluster data (with single star models) in \cite{gossage.etal:2021}.

\begin{table}[h!]
    \centering
    \caption{Adopted calibration parameters in the \cite{garraffo.etal:2018} braking model.}
    \begin{tabular}{||c|c|c||}
         Parameter &  This Work & \cite{garraffo.etal:2018} \\
         \hline
        $a$ & $0.03$ & $0.02$ \\
        $b$ & $0.5$ & $2.0$ \\
        $c$ & $3\times 10^{41}\ \rm{g}\ \rm{cm^{-2}}$ & $1\times 10^{41}\ \rm{g}\ \rm{cm^{-2}}$ \\
    \end{tabular}
    \label{tab:g18.params}
\end{table}

\subsubsection{\cite{van.ivanova:2019b} -- CARB braking}
We have also implemented the braking formalism utilised by \cite{van.ivanova:2019b} in their study of LMXBs. The authors follow a similar derivation to \cite{pavlovskii.ivanova:2016} (see also \citealt{van.ivanova:2019a}) and calculate the angular momentum lost through the Alfv\'en surface (as in \citealt{mestel.spruit:1987}), under the assumption of a radial magnetic field. They derive a form that is written as follows:

\begin{multline} \label{eq:carb19.jdot}
 \dot{J} = -\frac{2}{3}\dot{M}^{1/3}R^{14/3}(v_{\rm esc} + 2\Omega^2 R^2 / K_2^2)^{-2/3} \\
           \times \Omega_{\odot} B_{\odot}^{8/3} \left(\frac{\Omega}{\Omega_{\odot}}\right)^{11/3} \left(\frac{\tau_c}{\tau_{c,\odot}}\right)^{8/3}
\end{multline}
where $\dot{M}$ is the mass loss due to winds and $v_{\rm esc}$ is the escape speed. The solar magnetic field strength on its surface, $B_{\odot}$, is taken to be the average of about 1 G.The factor $(v_{\rm esc} + 2\Omega^2 R^2 / K_2^2)^{-2/3}$ takes into account the magnetocentrifugal acceleration of the stellar wind \citep{matt.etal:2012}. The constant $K_2$ encodes the threshold at which centrifugal effects become important in the wind. Here and in \citealt{van.ivanova:2019b}, a value of $K_2=0.07$ is adopted\footnote{$K_2$ appears in Eq. (6) of \cite{matt.etal:2012}, which can be rearranged as shown by \cite{reville.etal:2015}, appendix A, Eq. (A8). \cite{matt.etal:2012} found a value of $K_2=0.05$ best fit their MHD simulation results, which under rearrangement may be written as $K_2=0.07$, as done by \cite{reville.etal:2015}.}). \cite{van.ivanova:2019b} call this ``convection and rotation boosted'' (CARB) magnetic braking, in reference to its new dependence on both convection (via $\tau_c$) and greater dependence on rotation (via $\Omega$) in comparison to the Skumanich relation \citep{skumanich:1972}.

We also note that in contrast to the previously presented magnetic braking prescriptions, Eq. \ref{eq:carb19.jdot} contains dependence on the wind mass loss rate, $\dot{M}$, and is similar to the torque derived by \cite{matt.etal:2012}. This $\dot{M}$ dependence follows naturally from the solution to the angular momentum lost through the Alfv\'en surface, generally $\dot{J}_{mb}\propto \dot{M}R^2\Omega(r_A/R)^n$ \citep{mestel:1984,kawaler:1988}. However, \cite{pavlovskii.ivanova:2016} note that Skumanich-like scalings of $\dot{J}\propto B^2\Omega^3 R^4$ -- i.e., independent of $\dot{M}$ -- correspond to an assumption of isothermal winds in a radial field, as shown by \cite{mestel.spruit:1987}. Depending on the assumptions regarding e.g., magnetic field complexity and wind properties when parameterizing the Alfv\'en radius, $r_A$, this mass loss dependence may have different scalings or be non-existent. For instance, \cite{matt.etal:2012} retains a different dependence on $\dot{M}$ than Eq. \ref{eq:carb19.jdot}; this dependence is factored out in \cite{matt.etal:2015} through a phenomenological argument that relates $B$ and $\dot{M}$ to $R_{\rm o}$ and introduces the saturated and unsaturated regimes. Like the prescription of \cite{rappaport.verbunt.joss:1983}, Eq. \ref{eq:carb19.jdot} does not incorporate magnetic field saturation, in contrast to both \cite{matt.etal:2015} and \cite{garraffo.etal:2018}.

\begin{figure*}[!ht]
    \centering
    \includegraphics[width=\textwidth]{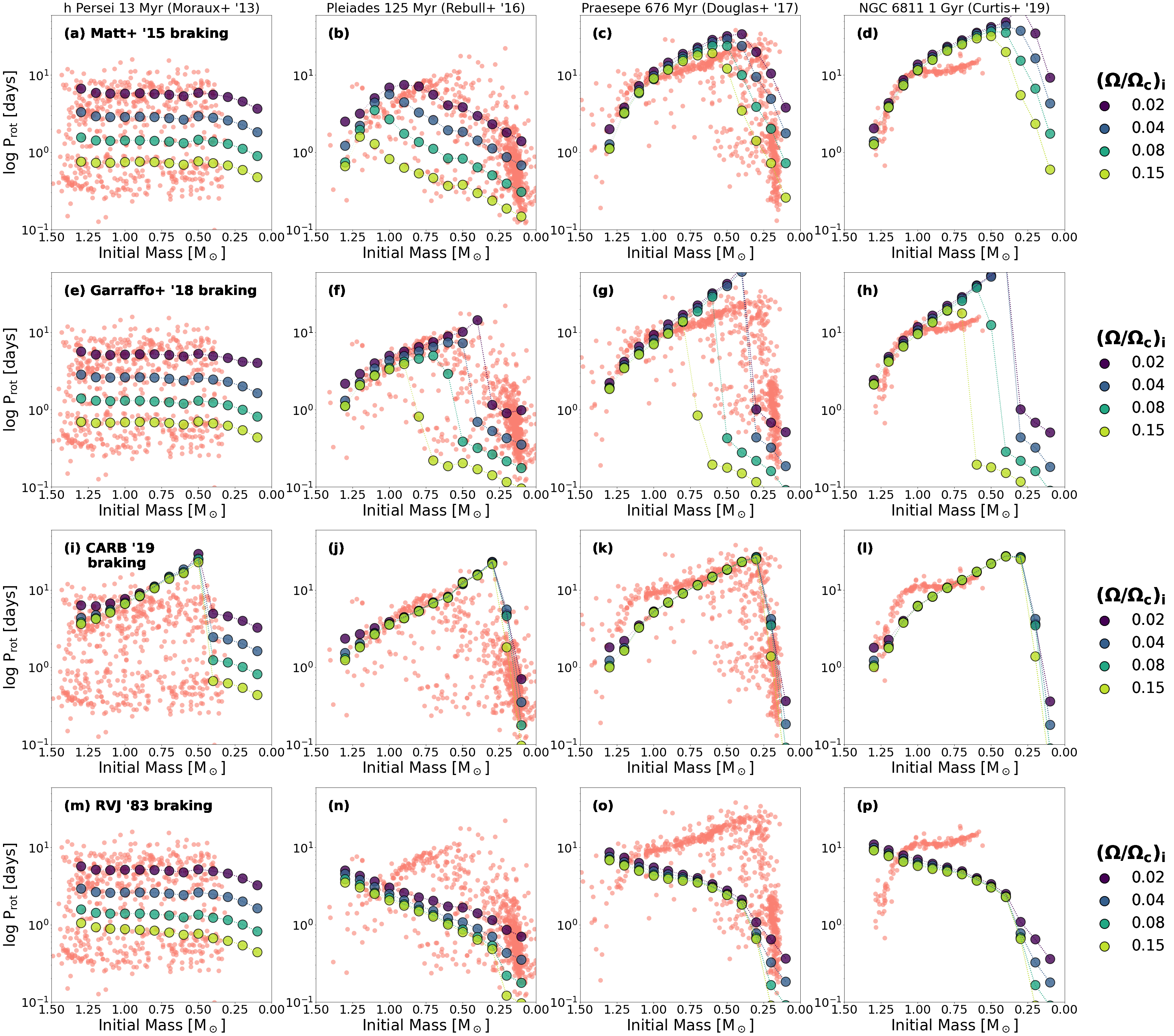}
    \caption{$\rm log\ P_{rot}$ is plotted against stellar mass. Advancing column-wise and rightwards, data from open clusters of increasing age are presented in red (cluster age, name, and reference appear in each column's title). Advancing row-wise, multi-colored points correspond to models following a specific braking prescription, annotated in the leftmost panel of each row; each color corresponds to a different value of $(\rm \Omega/\Omega_{crit})_i$. From top to bottom, the rows display single star models evolved under the \cite{matt.etal:2015}, \cite{garraffo.etal:2018}, \cite{van.ivanova:2019b} (CARB), and \cite{rappaport.verbunt.joss:1983} (RVJ83) magnetic braking prescriptions}
    \label{f:all.ocs}
\end{figure*}

\section{RESULTS} \label{sec:results}

We first present results from our single star analysis, and second from our binary star analysis in the following sections. Our results show that of the four magnetic braking prescriptions considered here, two (M15 and G18) can reproduce the evolution of rotation periods observed in open clusters. Of these two, G18 is capable of reproducing both single star rotation period and binary star LMXB period data. None of the prescriptions reproduce the stalled braking behavior seen in open cluster (single star) rotation period data (e.g., \citealt{curtis.etal:2020,hall.etal:2021}).

\subsection{Single Star Rotation Period Evolution}
We show the comparison between single star models evolved under each magnetic braking prescription and data from four representative open clusters in Fig. \ref{f:all.ocs}. The M15 \citep{matt.etal:2015} and G18 \citep{garraffo.etal:2018} prescriptions were both tested in \cite{gossage.etal:2021} against these data and their results remain similar to that study. The bottom two rows show results from the CARB \citep{van.ivanova:2019b} and RVJ83 \citep{rappaport.verbunt.joss:1983} prescriptions, which were primarily developed in the context of binary stellar evolution.

\subsubsection{\cite{matt.etal:2015} Torque}
Models evolved under M15 \citep{matt.etal:2015} trace the slow rotator branch well for masses of about $\rm 1.3 \geq M \geq 1\ M_{\odot}$ e.g., see Fig. \ref{f:all.ocs}, panels (a) - (c). At the age of the Pleiades (125 Myr) in Fig. \ref{f:all.ocs}, panel (b), data shows stars in the approximate range $\rm 0.9 \geq M \geq 0.5\ M_{\odot}$ have largely collected onto the slowly rotating branch, leaving a paucity of stars observed with rotation periods less than about 5 days. Under this prescription, our models tend to predict faster rotation rates in this mass range and at this age than what is observed. models towards the fully convective limit $\rm 0.4 M_{\odot} \geq M$ match the data fairly well, predicting that they reside on the rapidly rotating branch. At the age of the Praesepe (676 Myr), in Fig. \ref{f:all.ocs}, panel (c), models have now mostly settled onto the slowly rotating branch in the range $M \gtrsim \rm 0.6\ M_{\odot}$, but less massive models in the range $ \rm 0.5 \gtrsim M \gtrsim 0.3\ M_{\odot}$ tend to predict more rapid rotation than what is observed. This was discrepancy was also characterized by \cite{douglas.etal:2017}, and discussed by \cite{matt.etal:2015}, who attributed this behavior to deviations from solid body rotation, possibly related to a transition in internal angular momentum transport behavior near this age, and in this mass range. Lastly, by the age of the Praesepe, in Fig. \ref{f:all.ocs}, panel (c), and on towards that of NGC 6811 (1 Gyr), in Fig. \ref{f:all.ocs}, panel (d), models roughly within the mass range $\rm 1 > M > 0.6\ M_{\odot}$ tend to show slower rotation than what is observed, suggesting they are braking too efficiently between ~700 - 1000 Myr. \cite{curtis.etal:2019} have documented an observed epoch of stalled spin down within about 700 - 1000 Myr (at least) for a similar mass range, which may be related to internal angular momentum transport processes (e.g., see \citealt{spada.lanzafame:2020}), but is being actively researched.

\subsubsection{\cite{garraffo.etal:2018} Torque}
Models evolved under G18 \citep{garraffo.etal:2018} show a more rapid transition from fast to slow rotation, such that by the age of the Pleiades (125 Myr), in Fig. \ref{f:all.ocs}, panel (f), models in the range $\rm 1.3 \geq M \geq 0.9\ M_{\odot}$ have largely settled onto the slowly rotating branch, reflecting observations. Additionally, in the range $\rm 0.8 \geq M \geq 0.4\ M_{\odot}$, the models show a bifurcated behavior where slower models with $(\Omega/\Omega_c)_i \leq 0.04$ tend to reside on the slow branch, and those with $(\Omega/\Omega_c)_i \geq 0.08$ reside on the fast branch, with a paucity of models predicted to lie between the two branches, as observed. G18 models make a rapid transition from the rapid branch to the slow branch, so that by the age of the Praesepe (676 Myr), in Fig. \ref{f:all.ocs}, panel (g), the majority of the models with $M \gtrsim 0.3\ M_{\odot}$ reside on the slow branch (except for those models with initially more rapid rotation rates), reflecting data relatively well. This bifurcated behavior has only been replicated by the G18 prescription and the metastable dynamo model of \cite{brown:2014} so far. 

G18 also reproduces the overall trend where lower mass stars ($M \lesssim 0.3 M_{\odot}$), visible in Fig. \ref{f:all.ocs}, panels (f) and (g), tend rotate more quickly (similar to M15), as models under this prescription experience ``saturated'' behavior at lower $R_{\rm o}$ values; $R_{\rm o}$ is greater in lower mass stars with thicker convective envelopes (and greater $\tau_c$, see e.g., \citealt{cranmer.saar:2011}), causing them to remain saturated longer, and spin down at a slower rate. As noted in \cite{matt.etal:2015}, this trend may largely be attributed to the prescription's scaling with $\tau_{\rm conv}$ (as in, e.g., Eq. \ref{eq:g18.jdip}). Compared to M15, G18 predicts a faster transition between the saturated and unsaturated regimes. In the scenario of G18, ``saturation'' coincides with increased magnetic field complexity, which for these low mass rapid rotators would retain for a longer period of time. Like M15, G18 models experience the same issue of deviating from data beyond the age of the Praesepe and do not capture the stalled spin down observed by \cite{curtis.etal:2020,hall.etal:2021}.

The quicker transition between the fast and slow rotation branches with the G18 prescription is due to the exponential scaling found in Eq. \ref{eq:g18.qjsimple}. This scaling was derived in \cite{garraffo.drake.cohen:2016} by fitting equations to 3D MHD simulation results, and tested against several decomposed ZDI magnetograms (e.g., from \citealt{donati.landstreet:2009,marsden.etal:2011,alvarado-gomez.etal:2015,waite.etal:2015}). This exponential term depends on Eq. \ref{eq:g18.n} (with $n$ parameterizing magnetic complexity) which was also constructed to satisfy observational constraints (see Fig. 1 of \citealt{garraffo.etal:2018} and \citealt{gossage.etal:2021}) with magnetic complexity hindering spin down at both early and late stages (Sec. \ref{ssec:g18.form}). Incidentally, stalled spin down could possibly be captured through modification to Eq. \ref{eq:g18.n}, such that lower mass stars enter a phase of high magnetic field complexity (inefficient magnetic braking) at lower $R_{\rm o}$ than higher mass stars, stalling them at an earlier stage of their spin down than this equation currently predicts.

\begin{figure*}[!ht]
    \centering
    \includegraphics[width=\textwidth]{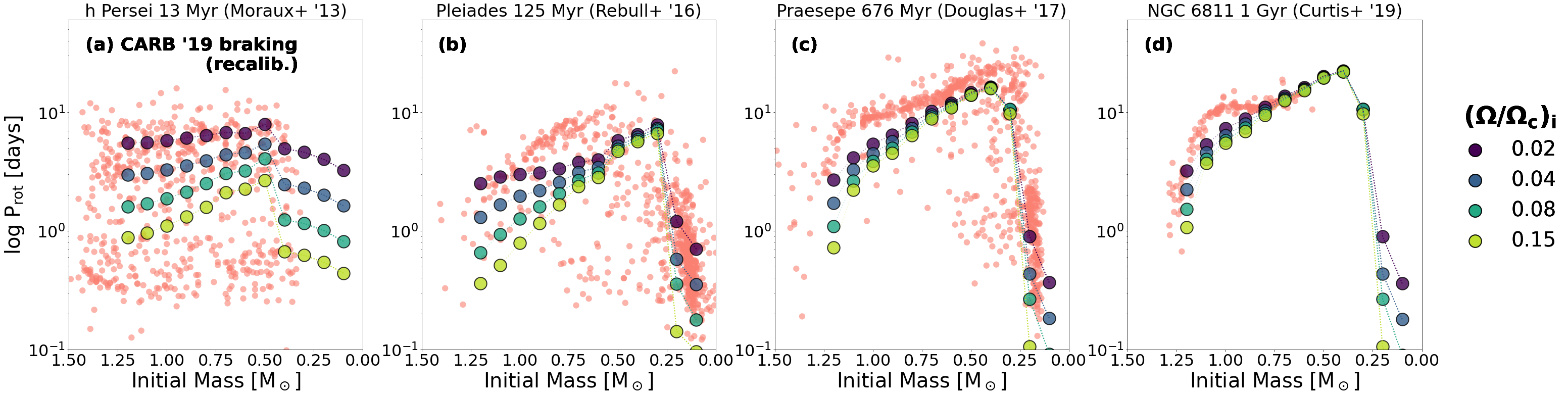}
    \caption{Similar to Fig. \ref{f:all.ocs}, $\rm log\ P_{rot}$ vs. stellar mass but only displaying models evolved under our recalibrated CARB \citep{van.ivanova:2019b} prescription, Sec. \ref{sssec:carb.ocs.results} (colored points; each color corresponding to a different value of $(\rm \Omega/\Omega_{crit})_i$). Red points are the data, with sources and open cluster ages listed in the title of each panel.}
    \label{f:carb.recal.ocs}
\end{figure*}

\subsubsection{\cite{van.ivanova:2019b} torque} \label{sssec:carb.ocs.results}
The CARB \citep{van.ivanova:2019b} prescription overestimates the spin down of (especially sub-solar mass) stars, such that models are already on the slow branch by the age of h Persei (13 Myr), in Fig. \ref{f:all.ocs}, panel (i). CARB braking causes an almost immediate spin down of stars post-disk locking (where we have assumed $\rm \tau_{DL}=3$ Myr) which does not represent the observed spin down behavior. However, CARB braking does replicate the observed periods of stars on the slow branch in the Pleiades, Fig. \ref{f:all.ocs}, panel (j), for masses in the range $\rm M \lesssim 0.5\ M_{\odot}$; masses in the range $\rm 0.5 \lesssim M \lesssim 0.3\ M_{\odot}$ appear to rotate more rapidly than observed, with models at all $(\Omega/\Omega_c)_i$ residing in the fairly sparse region of $\rm P_{rot} \gtrsim 10$ days. Fully convective models ($\rm 0.3\ M_{\odot} \lesssim M$) in the Pleiades, Fig. \ref{f:all.ocs}, panel (j), do not experience magnetic braking under this prescription, and are predicted to maintain rapid rotation rates that roughly match observations. At the ages of the Praesepe and NGC 6811 (676 and 1000 Myr), in Fig. \ref{f:all.ocs}, panels (k) and (l), models in the range $\rm 0.7 \geq M \geq 0.4\ M_{\odot}$ represent the data fairly well, but models in the range $\rm 1.3 \geq M \geq 0.8\ M_{\odot}$ tend to be braked too weakly, showing more rapid rotation than what is observed at these ages. Like M15 and G18, CARB struggles to replicate the observed stalled spin down in the age range 700 - 1000 Myr, as seen in Fig. \ref{f:all.ocs}, panels (k) and (l). Fully convective models in the mass range $\rm 0.3\ M_{\odot} \lesssim M$ roughly match observed values.

CARB magnetic braking does not incorporate saturated magnetic behavior in its formulation (unlike M15 and G18); however, it does possess scaling on $R_{\rm o}^{-1}$. The CARB prescription has a similar scaling of $R_{\rm o}^{-1}$ to the unsaturated, efficient braking regime of M15 (Eq. \ref{eq:m15.jdot.unsat}). Essentially these models show relatively efficient braking post-disk locking (in comparison to the other prescriptions) and arrive on the slow branch almost immediately. This is shown by the solid blue line in Fig. \ref{f:carb.recal}. CARB also does not replicate the rotation period of the Sun at the solar age, it predicts a present-day solar rotation rate of about 12 days. 

Fig. \ref{f:carb.recal} is a box and whiskers plot where boxes represent data from solar-like ($\rm 1.05 \gtrsim M \gtrsim 0.95\ M_{\odot}$) stars observed in variously aged open clusters annotated in the top of the plot. The vertical extent of each box gives a sense of the lower and upper quartile ranges of data (rotation periods) observed at a particular age for solar-like stars (from a particular open cluster). The red horizontal line in each box shows the median rotation period value. The ``whiskers'' extending beyond each box show the extent of the data out to roughly 3$\sigma$ from the median, while open circles represent outliers. Faint red dashed and/or dotted lines in Fig. \ref{f:carb.recal} are the three other braking prescriptions (M15, G18, and RVJ83), shown for reference. We only show the fastest rotation rate, $(\Omega/\Omega_c)_i = 0.15$, to simplify the comparison, but all rotation rates converge to the same value by the solar age.
\vspace{1cm}

\subsubsection{Recalibrated CARB braking}

\begin{figure*}[!t]
    \centering
    \includegraphics[width=0.75\textwidth]{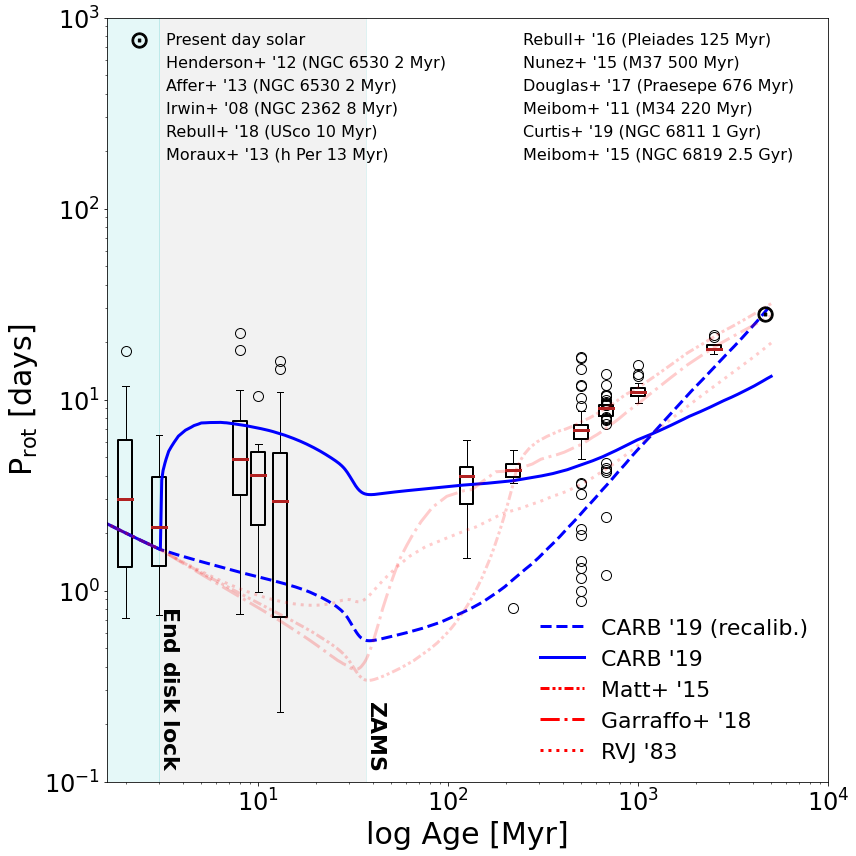}
    \caption{The (log) time evolution in Myr of $\rm log\ P_{rot}$; for a solar model ($\rm M =1\ M_{\odot}$, [Fe/H] = 0.0) under the CARB \citep{van.ivanova:2019b} magnetic braking prescription (solid blue) and our recalibration (dashed blue; Sec. \ref{sssec:carb.ocs.results}). We only show our fastest rotation rate, $(\rm \Omega/\Omega_{crit})_i=0.15$, for simplicity. Red lines of various styles show M15 \citep{matt.etal:2015}, G18 \citep{garraffo.etal:2018}, and RVJ83 \citep{rappaport.verbunt.joss:1983} for comparison. Box and whiskers (see text) show solar-like, $\rm 1.05 M_{\odot} > M > 0.95 M_{\odot}$, data from clusters whose name, age, and source are listed (top legend); open circles represent outliers. The solar value of $\rm P_{rot,\odot}$ adopted is 28 days, corresponding to the value $\rm \Omega_{\odot}=2.6\times 10^{-6}\ s^{-1}$ adopted in \cite{matt.etal:2015,matt.etal:2019}.}
    \label{f:carb.recal}
\end{figure*}

The dashed-blue line shown in Fig. \ref{f:carb.recal} is a recalibration of the CARB prescription, that we performed to try and match the solar rotation period at the solar age. In making our recalibration of Eq. \ref{eq:carb19.jdot.recalib}, we introduced a parameter similar to $p$, which is used to scale $R_{\rm o} = (\Omega\tau_c)^{-1}$ dependence in \cite{matt.etal:2015}; we call this new calibration parameter $p'$. Eq. \ref{eq:carb19.jdot} contains $R_{\rm o}^{-1}\sim \Omega\tau_c$ dependence due to a similar substitution and phenomenology used by \cite{matt.etal:2015} to re-parameterize the stellar magnetic field strength\footnote{However, \cite{matt.etal:2015} re-factor both the mass loss rate due to winds, $\dot{M}$ and stellar magnetic field strength $B$ into $R_{\rm o}^{-1}$, whereas \cite{van.ivanova:2019a} retain the dependence on $\dot{M}$ in their braking prescription.} (explained also in \citealt{van.ivanova:2019a} and references therein), and effectively has $p' \sim 2.7$. This value of $8/3$ (or about 2.7) follows from their chosen re-parameterization of the Alfv\'en radius under the radial magnetic field approximation. Introducing $p'$ and an overall scaling factor, $A$ to Eq. \ref{eq:carb19.jdot}, we have re-written it as

\begin{multline} \label{eq:carb19.jdot.recalib}
 \dot{J} = -A \frac{2}{3}\dot{M}^{1/3}R^{14/3}(v_{\rm esc} + 2\Omega^2 R^2 / K_2^2)^{-2/3} \\
           \times \Omega_{\odot} B_{\odot}^{p'} \left(\frac{\Omega}{\Omega_{\odot}}\right)^{p'+1} \left(\frac{\tau_c}{\tau_{c,\odot}}\right)^{p'}
\end{multline}
where now $p'$ may be adjusted to find a scaling of magnetic field strength with $R_{\rm o}^{-1}$ that satisfies the solar constraints. We tested several iterations, and found that $p'\approx1.3$ and $A\approx 10$ allowed models to replicate the solar rotation period at the solar age, and this weaker scaling on $R_{\rm o}^{-1}$ prevents the immediate spin down seen under Eq. \ref{eq:carb19.jdot}. 

Under this solar calibration, although $\rm P_{rot\odot}$ is now matched, models still struggle to replicate the open cluster period-mass data across time, as seen in Fig. \ref{f:carb.recal.ocs} (and as hinted in Fig. \ref{f:carb.recal} where the dashed-blue line tends to lie below the $\rm P_{rot}$ distributions of each cluster at later ages). In Fig. \ref{f:carb.recal.ocs}, panel (a), at the age of h Persei (13 Myr), models match data better, but still show greater spin down than what is observed roughly in the mass range $\rm 0.9 \geq M \geq 0.4\ M_{\odot}$. In panel (b) of Fig. \ref{f:carb.recal.ocs}, solar-like models display a range of rotation rates, whereas data shows stars in this mass range collapsed onto the slowly rotating sequence by the age of the Pleiades (125 Myr). Models in the mass range $\rm 0.9 \geq M \geq 0.4\ M_{\odot}$ tend to show more rapid rotation than what is observed, whereas previously models following the original CARB prescription replicated data reasonably well in roughly this mass range in Fig. \ref{f:all.ocs}, panel (j). At the age of the Praesepe (676 Myr), in Fig. \ref{f:carb.recal.ocs}, panel (c), all models above the fully convective limit (i.e., above roughly $\rm 0.3\ M_{\odot}$) are rotating more rapidly than observed, indicating this calibrated prescription is braking models too inefficiently. At the age of NGC 6811 (1 Gyr), in Fig. \ref{f:carb.recal.ocs}, panel (d), models may appear to roughly match data, but only because the data itself has undergone stalled spin down \cite{curtis.etal:2020,hall.etal:2021}, whereas the models have simply continued to spin down from panel (c), where they were already rotating too rapidly compared to the data. Overall, this recalibration does not simultaneously match solar and open cluster rotation period data; it is possible that further modifications to Eq. \ref{eq:carb19.jdot} could improve the fit to data. 

This is not a rigorous physical re-interpretation of Eq. \ref{eq:carb19.jdot}, but just a simple parameter test to reproduce $\rm P_{rot\odot}$. Physically, altering $p'$ as done here is similar to relaxing the radial magnetic field assumption, as this assumption determines the exponent $n$ in the general relation $\dot{J}_{mb}\propto \dot{M}R^2\Omega(r_A/R)^n$ (\citealt{mestel:1984,kawaler:1988}, where e.g., $n = 2$ for radial, and $n = 3/7$ for a dipole field) from which Eq. \ref{eq:carb19.jdot} follows. This assumption thus ultimately determines the exponent that goes on to scale magnetic field properties, which is characterized in Eqs. \ref{eq:carb19.jdot} and \ref{eq:carb19.jdot.recalib} through $R_{\rm o}^{-1}$ dependence.

A further point that this demonstrates is that the weaker $R_{\rm o}^{-1}$ dependence of Eq. \ref{eq:carb19.jdot.recalib}, which is more similar to a saturated (inefficient braking) state at early ages allows CARB to better replicate the data in h Persei at 13 Myr, as in Fig. \ref{f:carb.recal.ocs}, panel (a). Meanwhile stronger $R_{\rm o}^{-1}$ dependence, similar to an unsaturated (efficient braking) state, as in the original CARB prescription (Eq. \ref{eq:carb19.jdot}) better represents data at late ages, with a tight slow rotator sequence, i.e. in Fig. \ref{f:all.ocs}, panels (j) - (l). The effect of magnetic saturation incorporated in the M15 prescription captures a transition between these two states that allows models to experience inefficient magnetic braking at early ages and efficient braking later on. Transitioning between these two states gives M15 and G18 the flexibility to gradually shift stars between the fast and slow branch in a manner that replicates the spin down of stars in open clusters.

\subsubsection{\cite{rappaport.verbunt.joss:1983} Torque}
\label{sssec:rvj.ocs.results}
The RVJ83 \cite{rappaport.verbunt.joss:1983} prescription shows much slower spin down for sub-solar mass stars than what is observed, tending to predict faster rotation rates than exhibited by the data, especially in the approximate mass range $\rm 1 > M > 0.3\ M_{\odot}$ (125 - 1000 Myr), in Fig. \ref{f:all.ocs}, panels (n) - (p). RVJ83 braking (Eq. \ref{eq:rvj83.jdot}) scales with $MR^{\gamma}$ so that (for stars of a similar $\Omega$) lower mass stars experience a weaker torque and spin down slower. This is reflected in the overall trend shown by these models in the bottom row of Fig. \ref{f:all.ocs}, panels (m) - (p), where lower mass stars tend towards faster rotation periods. This prescription also does not simulate any ``saturated'' magnetic braking behavior (similar to CARB). Unlike the other prescriptions, it has no dependence on $R_{\rm o}$ (i.e., no $\tau_c$ scaling) and behaves similarly to always being in the saturated, inefficient braking regime -- almost opposite to the case of CARB. Likewise, this prescription does a fair job matching the spin evolution of stars residing on the fast branch (whereas CARB qualitatively matched the behavior of the slow branch). Solar-like rotation periods are roughly replicated by this prescription, which follows from its basing on the Skumanich law, although it under-predicts $\rm P_{rot\odot}$ at a value of about 18 days, as seen in Fig. \ref{f:carb.recal}.

\subsubsection{Summary of Single Star Results}
The success of the M15 and G18 prescriptions in replicating single star open cluster data is partially due to their inclusion of saturated (inefficient) and unsaturated (efficient) magnetic braking states. Both RVJ83 and CARB lack such a mechanism, and either always scale with $R_{\rm o}$ in the case of CARB (similar to always being unsaturated and efficiently braked), or never scale with $R_{\rm o}$ in the case of RVJ83 (similar to always being saturated and inefficiently braked). Consequently, CARB tends to brake stars immediately, never entering an inefficient braking phase as M15 and G18 do post-disk locking. RVJ83 shows weak braking at the onset, but remains weak and does not sufficiently slow models down in a manner resembling open cluster observations. Both CARB and RVJ83 assume that magnetic braking does not occur in fully convective stars, as data does suggest these low mass stars maintain rapid rotation rates and magnetic braking operates weakly on them.

Fig. \ref{f:carb.recal} also shows that CARB operates more strongly than the other prescriptions do for our solar model; \cite{van.ivanova:2019b} favored CARB as a stronger braking prescription (than \citealt{rappaport.verbunt.joss:1983} in particular) as it could replicate LMXB mass transfer rates (see \citealt{deng.etal:2021} as well who found similar, and). For our solar model, CARB braking is too efficient early on (in comparison to solar-like data presented in Fig. \ref{f:carb.recal}), reaching slow rotation rates quickly (during the pre-MS), at which point the braking rate slows. Thus, neither the observed solar-like young rapid rotators nor relatively slowly rotating Sun are reproduced. Under a recalibration (Sec. \ref{sssec:carb.ocs.results}) to match the solar rotation period at the solar age, the scaling on $R_{\rm o}$ was lessened (in essence relaxing the assumption of a radial magnetic field), weakening the initial braking that models experienced under CARB. The recalibrated CARB prescription better replicates the solar rotation period $\rm P_{rot\odot}$, but overall does not match the observed spin down stars through time in Fig. \ref{f:carb.recal.ocs}.

Of both M15 and G18, G18 is capable of replicating the paucity of stars observed in the region between the fast and slowly rotating branches of the data in Fig. \ref{f:all.ocs}, between the ages of the Pleiades and the Praesepe in panels (f) and (g). Whereas data shows that stars converge onto the slow sequence in the mass range of roughly $\rm M \geq 0.5\ M_{\odot}$ by the age of the Pleiades, M15 predicts that models have not converged, and instead still occupy the sparse region between the fast and slow branches by this age, in Fig. \ref{f:all.ocs}, panel (b). G18 does predict that models have converged to the slow sequence, and this transition happens rapidly due to exponential scaling of the braking efficiency (Eq. \ref{eq:g18.qjsimple}), leaving a sparsely populated region of stars between the fast and slowly rotating branches in Fig. \ref{f:all.ocs}, panel (f). This behavior persists to later ages. However, neither M15 nor G18 replicate the observed stalling of spin down and predict slower rotation rates overall for stars in this mass range at the age of the Praesepe and NGC 6811, i.e., panels (c), (d), (g), and (h) where the predicted slow branch of the models tends to lie above that observed in the data. Whether stalling may be attributed primarily to internal angular momentum transport (as modeled successfully by \citealt{spada.lanzafame:2020}), or whether other processes may be involved will need to be investigated in future work.

\subsection{LMXB Rotation Period Evolution}

\begin{figure*}[!ht]
    \centering
    \includegraphics[width=\textwidth]{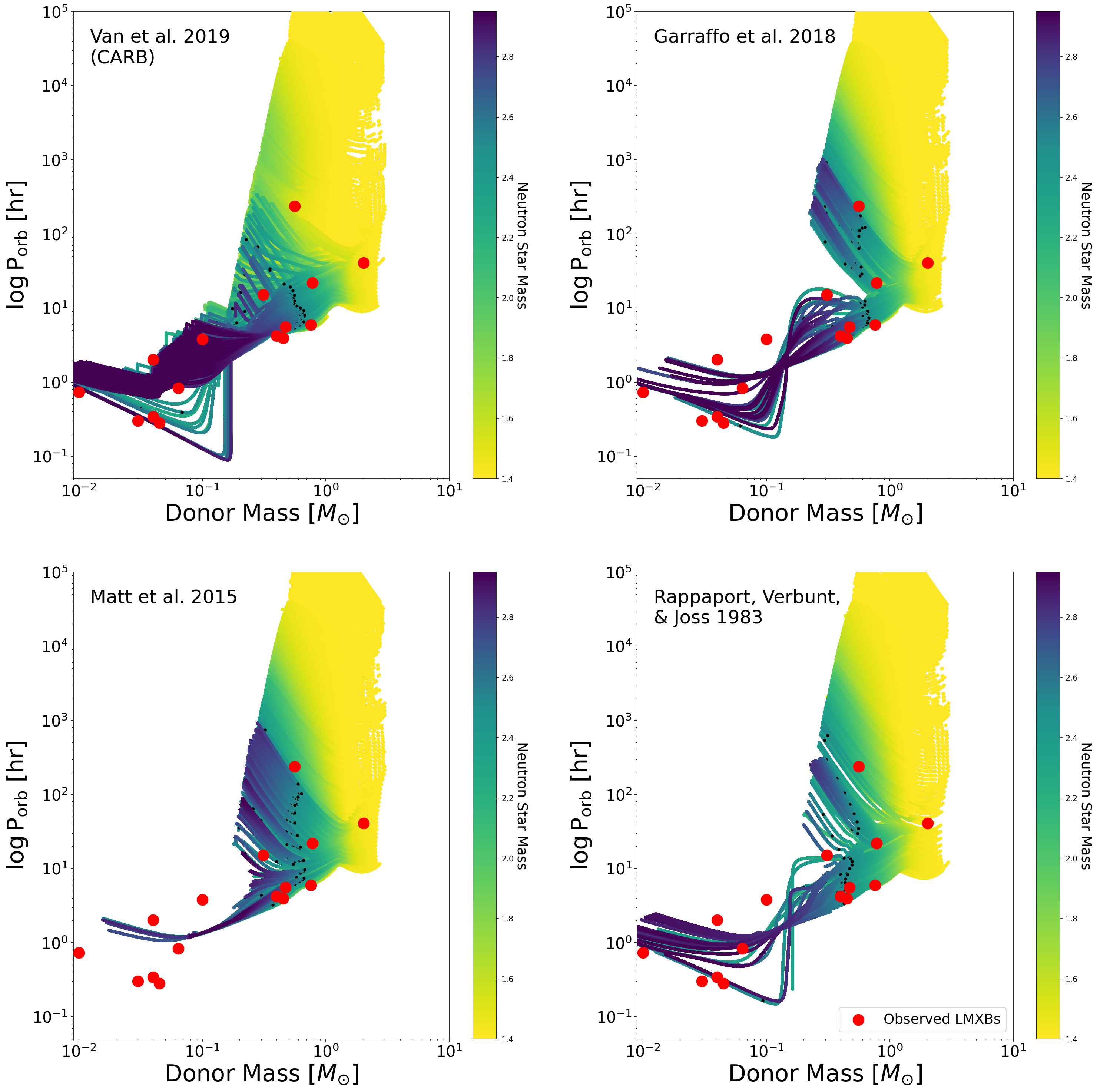}
    \caption{LMXB model log $P_{\rm orb}$ [hr] evolution vs. donor mass through mass transfer are displayed as the multi-colored lines. Each panel represents models evolved under a different magnetic braking prescription, annotated in the top left corner. The evolution proceeds from right to left as the donor star transfers mass to a $1.4\rm M_{\odot}$ companion representing a neutron star (modeled as a point mass). Color along each evolutionary track indicates the mass of the companion, and black dots indicate where the companion becomes a black hole according to the criterion used by \texttt{POSYDON} \citep{posydon.i:2022}. Red points are data from observed Milky Way LMXB systems, from the same data set used by \cite{van.ivanova:2019b}.}
    \label{f:grids.lmxbs}
\end{figure*}

In Fig. \ref{f:grids.lmxbs} we show the rotation period and donor mass evolution of our LMXB models during mass transfer versus several observed systems, for each magnetic braking prescription. \cite{van.ivanova:2019a} and \cite{van.ivanova:2019b} have shown comparisons for their CARB braking prescription, and the RVJ83 prescription; our results are similar to theirs in terms of replicating the observed LMXB rotation periods for CARB and RVJ83 braking. Fig. \ref{f:grids.lmxbs} shows that all four prescriptions qualitatively reproduce the bulk of the observations. However, M15 does not reproduce the observed UCXBs. In terms of \cite{lin.etal:2011}, these are the ``UC'' systems -- here we will use the label UCXB. The systems include 4U-0513-40, 4U-1543-624, and 4U-1850-087 in our data set, which are UCXB systems where the donors are thought to be helium or carbon-oxygen white dwarfs (see e.g., \citealt{heinke.etal:2013}  for more on these systems). These systems are only reproduced by our models that reach minimum $\rm P_{orb}$ less than about 30 minutes. We describe the conditions under which these systems are produced in the following section.

\subsubsection{UCXB Evolution and the Bifucation Period}
In general, the evolution of a UCXB begins with RLOF, where the orbit is shrunk through an interplay of both mass transfer and angular momentum loss. Systems that enter RLOF near the point of core hydrogen exhaustion typically achieve the most compact orbits, leaving the donor star as a helium white dwarf \citep{pylyser.savonije:1988, fedorova.ergma:1989, podsiadlowski.etal:2002, nelson.rappaport:2003, istrate.etal:2014}. There are several factors which influence whether or not a system will become a UCXB, but essentially, the orbit must shrink in order to produce small $P_{orb}$. \texttt{MESA} evolves the orbital separation ($\rm a_{sep}$) according to 

\begin{equation}\label{eq:a.sep}
    \rm{a_{sep}} = \frac{J_{\rm{orb}}^2}{G} \frac{(M_d + M_a)}{(M_d M_a)^2}
\end{equation}
\citep{paxton.etal:2015} where $J_{\rm orb}$ is the orbital angular momentum, and $M_d$ and $M_a$ are the donor and accretor masses, respectively. The orbital separation largely evolves in this case due to the competing effects of mass transfer and orbital angular momentum loss (of which magnetic braking is one source). During conservative mass transfer, the orbit shrinks if $M_d > M_a$, or expands if $M_d < M_a$ while $\dot{M}_d<0$ (see e.g., \citealt{verbunt:1993}). During angular momentum loss the orbit shrinks; the dominant source of angular momentum loss tends to be magnetic braking in our models, until orbits are very close and gravitational wave radiation becomes dominant. Whichever timescale is shorter (i.e., mass loss, $\tau_{ML}\sim M/|\dot{M}|$ or magnetic braking, $\tau_{J}\sim J_{\rm{orb}}/|\dot{J}_{\rm{orb}|}$) determines which effect dominates the orbital separation evolution. The evolutionary stage of the donor star at the onset of mass transfer plays a more nuanced role, and as discussed by e.g., \cite{rappaport.etal:1982,nelson.etal:1986} and \cite{nelson.rappaport:2003}, typically ultra short period UCXBs are produced in systems wherein the donor enters RLOF around the time of the terminal age MS (TAMS), before it becomes a subgiant.

\begin{figure}[!t]
    \centering
    \includegraphics[width=0.45\textwidth]{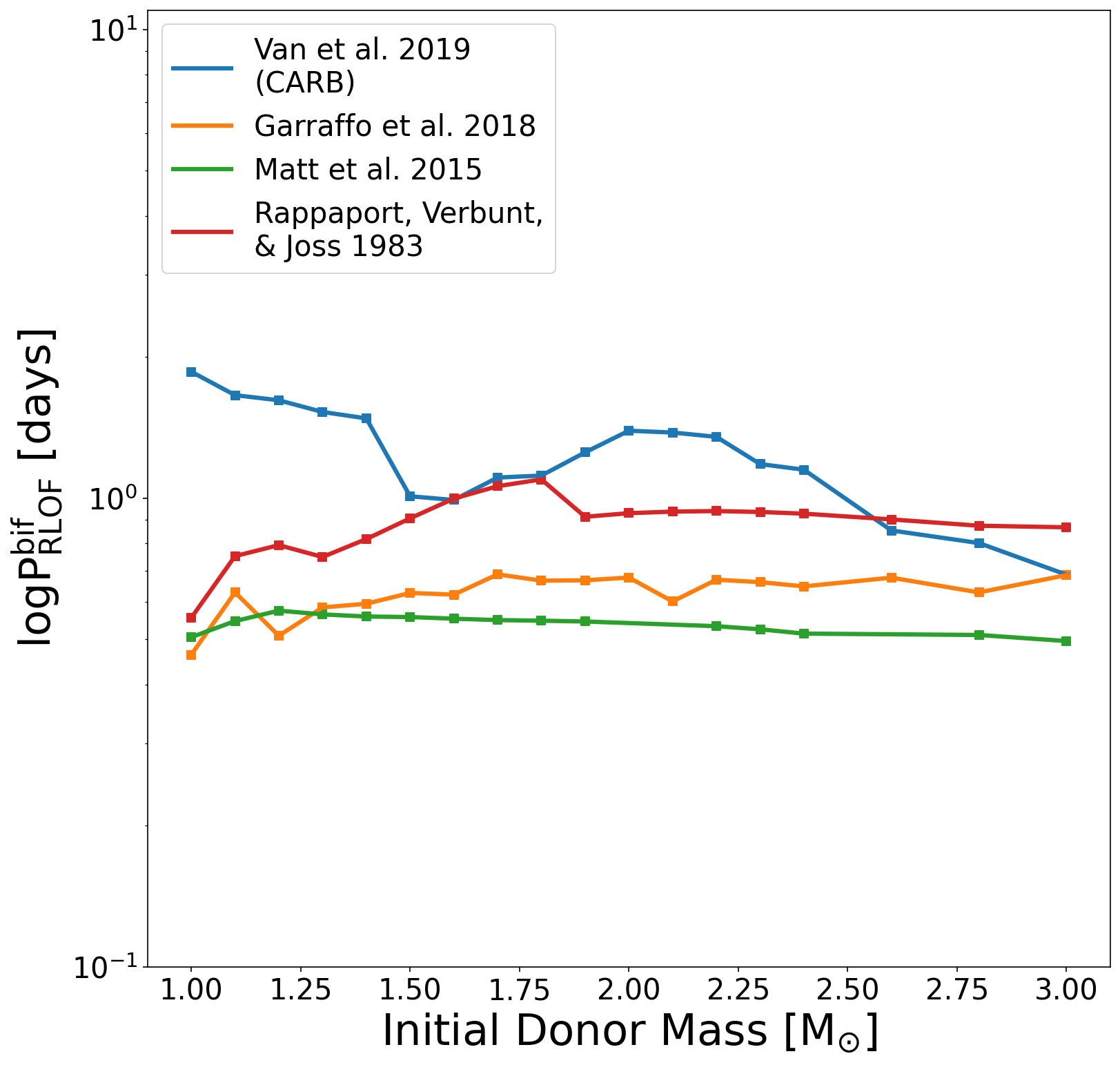}
    \caption{The bifurcation period at initial Roche lobe overflow, $\rm P^{bif}_{RLOF}$ vs. initial donor mass. Different colored lines correspond to models evolved with different magnetic braking prescriptions, labeled in the legend, and described in the text.}
    \label{f:grids.bifurcation}
\end{figure}

We find that our models do not produce these ultra short $\rm P_{orb} \lesssim 30^m$ UCXBs unless magnetic braking is sufficiently strong. One way to characterize the strength of a magnetic braking prescription is to examine its predicted bifurcation period. As defined by \cite{pylyser.savonije:1988,pylyser.savonije:1989}, the bifurcation period is the maximum orbital period below which binary systems converge to smaller $\rm P_{orb}$ than they began with, and above which systems diverge to larger $\rm P_{orb}$. We follow the definition used by  \cite{pylyser.savonije:1988,pylyser.savonije:1989} and plot the bifurcation period at initial RLOF, $\rm P_{RLOF}^{bif}$ for each initial donor mass grid point in Fig. \ref{f:grids.bifurcation} for each magnetic braking prescription.

Fig. \ref{f:grids.bifurcation} shows that $\rm P_{RLOF}^{bif}$ is sensitive to the treatment of magnetic braking, and also that overall M15 has smaller $\rm P_{RLOF}^{bif}$ than the other prescriptions. A consequence of this is that only relatively few UCXBs will form, which is also reflected in Fig. \ref{f:grids.lmxbs}, where M15 has fewer UCXBs compared to the other braking prescriptions. While the difference between bifurcations periods between M15 and G18 may seem small, it is enough to where G18 does reproduce $\rm P_{orb} \lesssim 30^m$ UCXBs, while M15 does not. Meanwhile, RVJ83 and CARB exhibit overall greater bifurcation periods, with CARB tending to have the greatest values, and greatest variety of systems that are able to form UCXB systems. We find that typically, systems just below the bifurcation period form the shortest period UCXBs, as in \cite{podsiadlowski.etal:2002}.

From our single star analysis, we found that CARB magnetic braking is highly efficient in extracting angular momentum, immediately slowing stars post-disk locking. We find that CARB magnetic braking becomes especially efficient with high mass loss, during RLOF. Owing to its greater strength, CARB shows the best performance in qualitatively reproducing the LMXB data in Fig. \ref{f:grids.lmxbs}. RVJ83 was found to be too weakly scaling overall in our single star analysis, yet it performs fairly well here; as may be seen in Fig. \ref{f:all.ocs}, RVJ83 actually slows systems with masses roughly greater than $1\rm\ M_{\odot}$ more strongly than the other prescriptions. As is the case here, where our donor masses are all initially greater than $1\rm\ M_{\odot}$, we find that RVJ83 actually produces overall stronger torques than the other prescriptions for the majority of the evolution. Lastly, as donor star models initially evolve in the unsaturated regime (roughly $R_{\rm o,sat} > 0.14$ in this case), and the G18 prescription produces a stronger torque than M15 does, it ultimately produces tighter orbits required to form UCXB systems. We find that M15 generally produces too weak of a torque to do the same.

It may be possible that a recalibration of M15 towards greater angular momentum loss could also allow it to reproduce $\rm P_{orb} \lesssim 30^m$ UCXBs, but this could also jeopardize its ability to fit the sequence of open cluster data (as in e.g., Fig. \ref{f:all.ocs} and Fig. \ref{f:carb.recal}). We have not exhausted all possible recalibrations of the M15 prescription, but note that a fairly simple rescaling of the law towards greater torque values would simply shift the curve (red, dashed, double dotted in Fig. \ref{f:carb.recal}) of M15 solar models systematically to greater $\rm P_{rot}$, disrupting its current fit to open cluster and solar data. A more complex adjustment of the mass scaling of M15 may be required to simultaneously match both the LMXB/UCXB and open cluster data.

\subsubsection{Summary of Binary Star Results}
While our results show that RVJ83, G18, and CARB are capable of reproducing LMXB/UCXB rotation periods (including the shortest period UCXBs) in Fig. \ref{f:grids.lmxbs}, the difficulties that RVJ83 and CARB have in reproducing the spin down of stars observed in open clusters (Fig. \ref{f:all.ocs}, bottom two rows) demonstrates inconsistent behavior across these two regimes. Although G18 appears to be a consistent description of magnetic braking across the binary and single star regimes, as seen in Fig. \ref{f:all.ocs}, panels (e) - (h) and the upper-right panel of Fig. \ref{f:grids.lmxbs}, we have not performed further tests as were done in \cite{van.ivanova:2019b} or, e.g, \cite{deng.etal:2021} to compare mass transfer rates or population densities of LMXB models. We also have not tested whether G18 can replicate a greater variety of UCXB systems as \cite{soethe.kepler:2021} show CARB is capable of. Testing whether G18 can also replicate these additional observational constraints is a goal of future work.

Additional evidence from binary star observations point towards saturation and possibly magnetic field complexity being an important ingredient of magnetic braking. \cite{garraffo.etal:2018b} showed that the CV period gap can naturally be explained upon considering variable magnetic field complexity. Fairly recently, \cite{pass.etal:2022} have also shown that data from wide M dwarf binaries favors a magnetic field morphology driven spin down. Also recently, \cite{el-badry.etal:2022} found that saturated magnetic braking laws like M15 and G18 were favored in comparisons to period distributions of CVs.

\section{DISCUSSION \& CONCLUSIONS} \label{sec:discussion.conclusions}

Our results suggest that of the four prescriptions examined, G18 performs best overall in the binary and single star regimes. However, it still does not replicate the stalled spin down behavior observed in open clusters \citep{curtis.etal:2019, curtis.etal:2020}. In this section we provide a discussion on the implication that magnetic saturation and magnetic complexity may be linked. As core-envelope coupling has recently been proposed as a promising model to reproduce stalled spin down behavior \citep{spada.lanzafame:2020}, we also briefly speculate on whether magnetic complexity is in contention with this theory. Lastly, we summarise our conclusions.

\subsection{Saturation and Magnetic Complexity}\label{ssec:sat.mag.disc}
Of the two prescriptions that incorporate saturation, M15 and G18, the latter reproduces the bimodal slow- and fast-branches observed in open clusters. The only other magnetic braking prescription capable of this is the phemenological metastable dynamo model (MDM) of \cite{brown:2014}, which incorporates elements of the semi-empirical model of \cite{barnes:2010} and \cite{barnes.kim:2010}. The MDM involves a random, spontaneous flip from a weakly coupled dynamo mode to a strongly coupled mode (i.e., coupling between the convection zone and radiative core). Stars are set to begin in the weakly coupled mode, creating the rapidly rotating (``C'') branch, and randomly flip to the strongly coupled mode where they settle onto to slowly rotating (``I'') branch in open cluster rotation period data. This spontaneous and random flip between coupling modes allows the MDM to replicate the morphology of these two branches.

Under the G18 prescription, angular momentum loss is exponentially damped (due to the factor \ref{eq:g18.qj}), leading to a relatively rapid transition between the rapidly and slowly rotating branches, and a sparsely populated gap between them, e.g., as seen in Fig. \ref{f:all.ocs}, panels (f) and (g). \cite{fritzewski.etal:2021} have also suggested that this transition between slowly and rapidly rotating branches appears to occur on relatively short timescales. The power-law scaling of the M15 prescription (Eqs. \ref{eq:m15.jdot.sat} and \ref{eq:m15.jdot.unsat}) produces a comparatively slower transition time. 

The exponential scaling of angular momentum loss (Eq. \ref{eq:g18.qj}) follows as a result of 3D MHD calculations carried out by \cite{garraffo.drake.cohen:2016}, however, it is less certain whether stars follow the prescribed evolution of magnetic field complexity as described by Eq. \ref{eq:g18.n}. This function was introduced in \cite{garraffo.etal:2018} and was motivated by two observational regimes. First, that ZDI magnetograms of young, rapidly rotating stars appear to store magnetic energy in higher order modes \citep{donati.landstreet:2009, marsden.etal:2011, waite.etal:2015}, suggesting high magnetic field complexity at low $R_{\rm o}$; a similar trend was also found via magnetogram analysis of late M dwarfs in \cite{garraffo.etal:2018b}. Second, that at late stages, solar like stars appeared to show slower rotation rates than expected \citep{vansaders.etal:2016,vansaders.pinsonneault.barbieri:2019}. The former coincides with the findings of e.g., \cite{noyes.etal:1984, pizzolato.etal:2003,wright.etal:2011,vidotto.etal:2014,wright.drake:2016,newton.etal:2017,wright.etal:2018,medina.etal:2020} that magnetic field activity also appears to saturate at rapid rotation rates, or generally at low $R_{\rm o}$, and the latter finding has been joined by the discovery of a stalled spin down observed between 700-1000 Myr \citep{curtis.etal:2019, curtis.etal:2020, hall.etal:2021}. While young, rapidly rotating stars may possess complex magnetic fields (as also mentioned by \citealt{matt.pudritz:2008a} based on ZDI results from \citealt{johns-krull:2007}), there is less evidence that older, slowly rotating stars may possess complex magnetic fields as well, and whether magnetic field complexity could also be responsible for the recently observed late-stage stalling of stellar spin down.

Furthermore, results from \cite{finley.matt:2017,finley.matt:2018} have shown via 2.5D MHD simulations that the Alfv\'en surface of a star may be primarily determined by the star's dominant lowest order mode in mixed field topologies (i.e., combined di-, quadru-, and/or octupolar modes), highlighting that higher order modes may tend to play a diminished role. Studies by \cite{see.etal:2019} and \cite{see.etal:2020}, working from the model of \cite{finley.matt:2018}, showed that higher order modes must be considered when a star's mass loss rate is above a critical level; in estimating the mass loss rates of numerous stars utilizing ZDI maps, they found that relatively few met this criteria, and those that did occurred at larger Rossby numbers (somewhat contrary to the trend proposed via Eq. \ref{eq:g18.n}). However, it is important to note that considerable differences exist between the simulations of \cite{garraffo.drake.cohen:2015,garraffo.drake.cohen:2016} and those of \cite{finley.matt:2017,finley.matt:2018}, upon which the results of \cite{see.etal:2019,see.etal:2020} are based.

The 3D MHD simulations by \cite{garraffo.drake.cohen:2015,garraffo.drake.cohen:2016} utilize the Alfv\'en Wave Solar Model (AWSoM; \citealt{vanderHolst.etal:2014}), which self consistently evolves the coronal heating and wind acceleration via the MHD calculations, which have been verified against solar data (e.g., \citealt{sachdeva.etal:2021}). The 2.5D MHD simulations by \cite{finley.matt:2017,finley.matt:2018} take a common approach, which is to use a polytropic wind model. As discussed by \cite{pantolmos.etal:2017} and \cite{finley.matt.see:2018} the polytropic wind model contains some limitations in its ability to model the winds of real stars, as is also discussed by \cite{cohen:2017} in comparison to e.g., the AWSoM model. It is unclear if solely the difference in wind model accounts for all discrepancies between these models, or in any case what the precise source(s) for their differences may be; this should be resolved in future works. Additional insights from ZDI map information will also aid in understanding the evolution of magnetic field complexity.

\subsection{Stalled Spin Down Between 700 - 1000 Myr}
None of the magnetic braking prescriptions under consideration here can resolve the outstanding issue of stalled spin down observed recently by \cite{curtis.etal:2019, curtis.etal:2020, hall.etal:2021} in 700 - 1000 Myr open clusters. The only prescription that is capable of this so far is that of \cite{spada.lanzafame:2020} in the scenario of core-envelope coupling. This model suggests that the radiative core and convective envelope may become weakly coupled, leading to inefficient angular momentum transport. The rapidly rotating core could then more or less slowly replenish angular momentum lost in the decoupled envelope due to magnetised winds, helping a star maintain a more rapid rotation rate (thus stalling its spin down) over some period of time. As lower mass stars have increasingly deep convection zones, eventually becoming fully convective, it seems reasonable that the core and envelope could have some varying degree of coupling dependent on mass, ultimately affecting the efficiency of internal angular momentum transport.

Conservatively, the model of \cite{spada.lanzafame:2020} may be able to explain the stalled spin down of stars without the need to invoke late stage magnetic field complexity (as in Eq. \ref{eq:g18.n}). However, presumably the degree of core-envelope coupling could have some impact on the stellar dynamo responsible for generating the magnetic field \citep{barnes:2003}, possibly impacting its geometry. The degree to which this may occur will likely remain speculation until further ZDI observations are gathered and analyzed to better understand the evolution of stellar magnetic field geometries. Nonetheless, a possibility remains that the two could be interrelated. Recently \cite{dungee.etal:2022} found that core-envelope decoupling models were broadly consistent with stalled spin down behavior up to the 4 Gyr old cluster M67, although the model of \cite{spada.lanzafame:2020} predicted slightly faster rotation periods than what is observed at that age. G18 appears to be a physically motivated model that can explain the bimodal nature of the fast and slow sequences observed in open cluster rotation period data. As discussed by \cite{brun.browning:2017}, the interplay between rotation, convection, magnetism is complex and it is likely that alterations in the changes in the dynamo could affect magnetic field geometry at least in some way.

\subsection{Conclusions and Future Work}
We have analyzed 1D single and binary stellar models (\texttt{MESA r11701}; \citealt{paxton.etal:2011,paxton.etal:2013,paxton.etal:2015,paxton.etal:2018,paxton.etal:2019}) in comparison to rotation period and mass data from open clusters and LMXBs. These 1D stellar models were evolved under four different magnetic braking prescriptions commonly discussed in recent literature: RVJ83 \citep{rappaport.verbunt.joss:1983}, CARB \citep{van.ivanova:2019b}, M15 \citep{matt.etal:2015}, G18 \citep{garraffo.etal:2018}. Prescriptions from RVJ83 and CARB were primarily developed in the context of, and are commonly used in studies of binary stellar evolution. Meanwhile, M15 and G18 have primarily been discussed and applied in the context of single star evolution.

Overall, the G18 prescription is capable of successfully modeling stellar spin evolution in both the single and binary star regimes, whereas other prescriptions have difficulties in one regime or the other. Thus far, the G18 prescription and the metastable dynamo \citep{brown:2014} are the only models demonstrated to be capable of replicating the fast and slow branches in open cluster data, with a sparsely populated gap between them. In the case of G18, this is due to its more rapid transition, owing to its exponential scaling (derived in \citealt{garraffo.drake.cohen:2016}), versus e.g. power law scaling used in other prescriptions. Physically, G18 links this behavior to magnetic field complexity (see also the discussion in Sec. \ref{ssec:sat.mag.disc}). In addition, the G18 prescription qualitatively replicates the observed spin and donor masses of several LMXB systems observed in the Milky Way, including UCXB systems.

We also find that the braking prescriptions commonly used in binary stellar evolution studies -- i.e., RVJ83 and CARB in this study -- do not replicate the spin down behavior observed in open clusters. A major reason appears to be the lack of a mechanism to transition between so-called ``saturated'' and ``unsaturated'' modes, as is accounted for in prescriptions from M15 and G18. This is discussed further in Sec. \ref{sssec:carb.ocs.results} and Sec. \ref{sssec:rvj.ocs.results}, where it is noted that essentially CARB braking always scales with $R_o$ as if it is in the unsaturated mode, and RVJ83 as if it is always in the saturated mode. The saturated mode scaling moreso replicates the fast rotator branch, while the unsaturated mode moreso replicates the slow rotator branch; this was also found and demonstrated by \cite{matt.etal:2015}.

Altogether, this suggests that future magnetic braking prescriptions should incorporate magnetic saturation in some way, as in M15 and G18, to replicate the fast and slow rotator branches, and the transition between them that is observed in open cluster data. The transition between these branches appears to happen on relatively short timescales, and can be matched roughly by the G18 prescription (see alternatively \citealt{brown:2014}). The RVJ83 and CARB prescriptions do not transition between these two branches, and virtually collapse onto one or the other. M15 does transition, but on longer timescales, leaving more stars in the ``gap'' between the branches than what may be observed. Additional investigations into the magnetic field geometry of all stars, and especially M dwarfs should be sought to confirm whether stars actually spin down as observed due to complex magnetic field geometries.

None of the prescriptions implemented in this study capture the stalled spin down observed in 700 - 1000 Myr old open clusters \citep{curtis.etal:2020,hall.etal:2021}; the only current prescription demonstrated to do so is the core-envelope coupling model of \cite{spada.lanzafame:2020}. In future work we aim to explore both the effects of higher order magnetic fields on spin down and the core-envelope coupling mechanism, as data currently suggests both effects may be crucial. We also aim to utilize further observational constraints from both the single and binary star regimes to continue work towards a consensus magnetic braking prescription. In doing so, we work towards building a consistent picture of the processes that drive stellar spin evolution in both regimes.

\section{Data Availability}
The \texttt{MESA} inlists, source code, and output for our models may be found on \texttt{Zenodo}: \dataset[10.5281/zenodo.7679543]{\doi{10.5281/zenodo.7679543}} with further explanations.

\acknowledgments
We thank the anonymous referee for their helpful comments in improving this manuscript. SG acknowledges funding from Northwestern University through a CIERA Postdoctoral Fellowship. VK was partially supported through a CIFAR Senior Fellowship, a Guggenheim Fellowship, and the Gordon and Betty Moore Foundation through grant GBMF8477. MS also acknowledges support through the  GBMF8477 grant (PI Kalogera). This research was supported in part through the computational resources and staff contributions provided for the Quest high performance computing facility at Northwestern University which is jointly supported by the Office of the Provost, the Office for Research, and Northwestern University Information Technology.

%


\software{This work has used \texttt{MESA r11701} \citep{paxton.etal:2011,paxton.etal:2013,paxton.etal:2015, paxton.etal:2018, paxton.etal:2019} and the Python modules \texttt{numpy} \citep{harris.etal:2020} and \texttt{matplotlib} \citep{hunter:2007}.}





\bibliographystyle{aasjournal}



\end{document}